\documentclass[amsmath,showpacs,nofootinbib,11pt,superscriptaddress]{revtex4-2}
\usepackage{graphicx}
\usepackage{dcolumn}
\usepackage{bm}
\usepackage{color} 
\usepackage{slashed}
\usepackage{amsfonts}
\usepackage{caption}
\usepackage{subcaption}
\begin{document}
\newcommand{\hs}{\hspace*{0.5cm}}
\newcommand{\vs}{\vspace*{0.5cm}}
\newcommand{\be}{\begin{equation}}
\newcommand{\ee}{\end{equation}}
\newcommand{\bea}{\begin{eqnarray}}
\newcommand{\eea}{\end{eqnarray}}
\newcommand{\ben}{\begin{enumerate}}
\newcommand{\een}{\end{enumerate}}
\newcommand{\bde}{\begin{widetext}}
\newcommand{\ede}{\end{widetext}}
\newcommand{\nn}{\nonumber}
\newcommand{\crn}{\nonumber \\}
\newcommand{\Tr}{\mathrm{Tr}}
\newcommand{\non}{\nonumber}
\newcommand{\noi}{\noindent}
\newcommand{\al}{\alpha}
\newcommand{\la}{\lambda}
\newcommand{\bet}{\beta}
\newcommand{\ga}{\gamma}
\newcommand{\va}{\varphi}
\newcommand{\om}{\omega}
\newcommand{\pa}{\partial}
\newcommand{\+}{\dagger}
\newcommand{\fr}{\frac}
\newcommand{\bc}{\begin{center}}
\newcommand{\ec}{\end{center}}
\newcommand{\Ga}{\Gamma}
\newcommand{\de}{\delta}
\newcommand{\De}{\Delta}
\newcommand{\ep}{\epsilon}
\newcommand{\varep}{\varepsilon}
\newcommand{\ka}{\kappa}
\newcommand{\La}{\Lambda}
\newcommand{\si}{\sigma}
\newcommand{\Si}{\Sigma}
\newcommand{\ta}{\tau}
\newcommand{\up}{\upsilon}
\newcommand{\Up}{\Upsilon}
\newcommand{\ze}{\zeta}
\newcommand{\ps}{\psi}
\newcommand{\Ps}{\Psi}
\newcommand{\ph}{\phi}
\newcommand{\vph}{\varphi}
\newcommand{\Ph}{\Phi}
\newcommand{\Om}{\Omega}
\newcommand{\AdrHEPC}{Phenikaa Institute for Advanced Study, Phenikaa University, Yen Nghia, Ha Dong, Hanoi 12116, Vietnam}
\newcommand{\AdrIOP}{Institute of Physics, Vietnam Academy of Science and Technology, 10 Dao Tan, Ba Dinh, Hanoi 100000, Vietnam\vspace{0.2cm}}

\title{Scoto-seesaw model implied by flavor-dependent Abelian gauge charge} 

\author{Duong Van Loi}
\email{loi.duongvan@phenikaa-uni.edu.vn}
\affiliation{\AdrHEPC}
\author{N. T. Duy}
\email{ntduy@iop.vast.vn}
\affiliation{\AdrIOP}
\author{Cao H. Nam}
\email{nam.caohoang@phenikaa-uni.edu.vn}
\affiliation{\AdrHEPC}
\author{Phung Van Dong}
\email{dong.phungvan@phenikaa-uni.edu.vn (corresponding author)}
\affiliation{\AdrHEPC}

\date{\today}

\begin{abstract}
Assuming fundamental fermions possess a new Abelian gauge charge that depends on flavors of both quark and lepton, we obtain a simple extension of the Standard Model, which reveals some new physics insights. The new gauge charge anomaly cancellation not only explains the existence of just three fermion generations as observed but also requires the presence of a unique right-handed neutrino $\nu_R$ with a non-zero new gauge charge. Further, the new gauge charge breaking supplies a residual matter parity, under which the fundamental fermions and $\nu_R$ are even, whereas a right-handed neutrino $N_R$ without the new charge is odd. Consequently, light neutrino masses in our model are generated from the tree-level type-I seesaw mechanism induced by $\nu_R$ and from the one-loop scotogenic contribution accommodated by potential dark matter candidates, $N_R$ and dark scalars, odd under the matter parity. We examine new physics phenomena related to the additional gauge boson, which could be observed at colliders. We analyze the constraints imposed on our model by current experimental limits on neutrino masses, neutral meson oscillations, $B$-meson decays, and charged lepton flavor violating processes. We also investigate the potential dark matter candidates by considering relic density and direct detection.
\end{abstract}

\maketitle

\section{Introduction}
The Standard Model (SM) of particle physics has been remarkably successful in describing the fundamental particles and their interactions. However, the discovery of neutrino oscillations \cite {RevModPhys.88.030501, RevModPhys.88.030502} and the observation of a dark matter (DM) relic density that makes up most of the mass of galaxies and galaxy clusters \cite{WMAP:2012nax,Planck:2018vyg} thus call for physics beyond the SM. Additionally, the SM cannot explain the existence of just three fermion generations, as observed, and several flavor puzzles in both quark and lepton sectors \cite{ParticleDataGroup:2022pth}.

On the other hand, the canonical seesaw mechanism is popularly accepted for generating appropriate small neutrino masses \cite{Minkowski:1977sc,GellMann:1980vs,Yanagida:1979as,Glashow:1979nm,Mohapatra:1979ia,Mohapatra:1980yp,Lazarides:1980nt,Schechter:1980gr,Schechter:1981cv}. This is achieved at the tree level by introducing three heavy Majorana right-handed neutrino singlets $\nu_{1,2,3R}$ into the SM. However, in its simple form, this mechanism does not naturally address the issue of DM unless some DM stability condition or parameter finetuning is {\it ad hoc} imposed. Although the neutrino mass and DM may be uncorrelated, it is worth exploring scenarios where both issues can be addressed in the same solution. Precisely, this happens in the scotogenic mechanism, where light neutrino masses arise at the quantum level via loops involving dark messengers that may also be suitable for DM particles \cite{Ma:2006km}. The most economical version of the scotogenic mechanism requires a couple of fermionic singlets and an inert scalar doublet, which are odd under an assumed symmetry, $\mathbb{Z}_2$. The assumed symmetry stabilizes the lightest odd particle and provides either a fermion or scalar DM particle. Moreover, nonzero neutrino masses can also be generated through a hybrid mass mechanism comprising seesaw and radiative mass mechanisms, known as scoto-seesaw \cite{Kubo:2006rm,Rojas:2018wym}.

It is well known that the SM is based on the gauge symmetry $SU(3)_C\otimes SU(2)_L\otimes U(1)_Y$. The first factor, $SU(3)_C$, is the ordinary QCD symmetry, while the second factor, $SU(2)_L$, is the symmetry of weak isospin $T_i$ ($i=1,2,3$). The last factor, $U(1)_Y$, is called the weak hypercharge symmetry, which ensures the algebraic closure between electric charge $Q$ and the weak isospin $T_i$. Further, the value of hypercharge $Y$ is chosen to describe observed electric charges via $Q=T_3+Y$. Notice that the charges $Q$, $T_i$, and $Y$ are universal for every flavor of neutrinos, charged leptons, up-type quarks, and down-type quarks. An interesting question relating to these charges is whether their universality causes the SM to be unable to address the issues. The present work does not directly answer such a question. Instead, we look for a new Abelian gauge charge depending on flavors of both quark and lepton, which naturally solves the issues.

To achieve this aim, we extend the gauge symmetry of SM by including a new Abelian gauge symmetry, called $U(1)_X$, such that the new Abelian gauge charge, $X$, is dependent on the flavor (through a generation index $a$) of both quarks (via the baryon number $B$) and leptons (via the lepton number $L$). Additionally, we require that quark generations, as well as lepton generations, carry the new gauge charges either the same or opposite in sign to reduce degrees of freedom in the model for simplicity. We also suggest that the new gauge charge of the third quark (the first lepton) generation should differ from that of the first and second quark (the second and third lepton) generations. Thus, we find the expression for the new gauge charge in the following form,
\be X=3z[Bi^{a^2(a-1)}+Li^{a(a-1)}],\label{chargeX} \ee
where $z$ is an arbitrary nonzero parameter and $i$ is the imaginary unit.\footnote{In recent work, we considered such a charge for flavor questions but not DM \cite{VanLoi:2023kgl}.} Notice that $X$ is Hermitian because $i^{a(a-1)} = (-1)^{a(a-1)/2}$ is always real. Interestingly, the anomaly cancellation of the new gauge charge explains the existence of just three fermion generations, as observed. It also requires the presence of a unique right-handed neutrino $\nu_R$ with a non-zero new gauge charge. Further, the new gauge charge breaking supplies a residual matter parity, for which the SM fermions and $\nu_R$ are even. In contrast, a right-handed neutrino $N_R$ without the new gauge charge is odd. These naturally motivate the generation of light neutrino masses from the tree-level type-I seesaw mechanism induced by $\nu_R$ and from the one-loop scotogenic mechanism accommodated by $N_R$ and dark scalars odd under the matter parity. We investigate the resulting model with a minimal scalar content in detail and analyze the constraints imposed on the model by current experimental limits on neutrino masses, neutral meson oscillations, $B$-meson decays, and charged lepton flavor violating (cLFV) processes. We also examine new physics (NP) phenomena related to the additional gauge boson, which could be observed at colliders, and investigate the DM candidates by considering relic density and direct detection. 

This paper is organized as follows. In Sect. \ref{model}, we present our model and discuss its essential aspects, such as gauge symmetry, particle content, charge assignment, and residual matter parity. We also examine the mass spectrum of fermions, scalars, and gauge bosons, then determine the couplings of physical gauge fields with fermions. Collider bounds from the additional gauge boson are discussed in Sect. \ref{collider}. In Sect. \ref{flavor}, we analyze the constraints imposed on the model by current experimental limits on neutral meson mixings, $B$-meson decays, and cLFV processes. The DM candidates are investigated by considering relic density and direct detection as in Sect. \ref{dm}. Finally, we summarize our results and conclude this work in Sect. \ref{conclusion}.

\section{\label{model}The model}
\subsection{Gauge symmetry and fermion content}
As mentioned above, our model is based on gauge symmetry,
\be SU(3)_C\otimes SU(2)_L\otimes U(1)_Y\otimes U(1)_X,\label{gaugesymmetry} \ee
in which the first three factors are precisely the gauge symmetry of SM, and the last factor is the additional Abelian gauge symmetry with the charge $X$ to be determined in Eq. (\ref{chargeX}). The SM fermion multiples possess new gauge charges dependent on generations via an index $a$, as shown in Table \ref{tab1}. 
\begin{table}[h]
\bc
\begin{tabular}{l|cccccccc}
\hline\hline
Multiplets & $l_{aL}=(\nu_{aL},e_{aL})^T$ & $e_{aR}$ & $q_{aL}=(u_{aL},d_{aL})^T$ & $u_{aR}$ & $d_{aR}$\\ \hline
$U(1)_X$ & $3zi^{a(a-1)}$  & $3zi^{a(a-1)}$ & $zi^{a^2(a-1)}$ & $zi^{a^2(a-1)}$ & $zi^{a^2(a-1)}$\\
\hline\hline
\end{tabular}
\caption[]{\label{tab1}$U(1)_X$ charges of SM fermions, where $a$ is a generation index.}
\ec
\end{table}

Interestingly, the new gauge charges of fermion generations are periodic in $a$ with a period of $4$. Indeed, with $a=1,2,3,4,5,6,7,8,\cdots$, then $X=z,z,-z,z,z,z,-z,z,\cdots$ for the quark generations and $X=3z,-3z,-3z,3z,3z,-3z,-3z,3z,\cdots$ for the lepton generations. Hence, it is convenient to express the number of fermion generations $N_f$ as $N_f=4m-n$, where $m=1,2,3,\cdots$, and $n=0,1,2,3$. Considering the gauge anomaly $[SU(2)_L]^2U(1)_X$, we obtain
\be[SU(2)_L]^2U(1)_X \sim \sum_{\mathrm{doublets}}X_{f_L}=\left\{\begin{array}{ll}6z(m-1) &\text{ if }n=1\\
6zm&\text{ if }n=0,2,3\end{array}.\right.\ee
This anomaly is canceled if and only if $m=n=1$. Thus, the number of fermion generations is precisely three, $N_f=3$, as observed.  
 
With $N_f=3$ and the fermion content as in Table \ref{tab1}, two anomalies, 
\bea
\left[\text{Gravity}\right]^2U(1)_X &\sim& \sum_{\mathrm{fermions}}(X_{f_L}-X_{f_R})=-3z,\\
\left[U(1)_X\right]^3 &\sim&\sum_{\mathrm{fermions}}(X^3_{f_L}-X^3_{f_R})=-27z^3,\eea
are not canceled yet. To cancel these anomalies as well as generate appropriate neutrino masses (see below), we introduce two right-handed neutrinos,
\be \nu_R\sim ({\bf 1}, {\bf 1}, 0,-3z),\hs N_R\sim ({\bf 1}, {\bf 1}, 0,0),\label{neuR}\ee into the theory as fundamental constituents. The fermion content of our model and their quantum numbers are displayed in Table \ref{tab2}, where we conveniently define two kinds of generation indices, like $\al,\bet$ run over $1,2$ for the first two quark generations, while $x,y$ run over $2,3$ for the last two lepton generations; generically, $a,b$ run over $1,2,3$ according to $N_f=3$.
\begin{table}[h]
\bc
\begin{tabular}{lccccc|lccccccccc}
\hline\hline
Multiplets & $SU(3)_C$ & $SU(2)_L$ & $U(1)_Y$ & $U(1)_X$ & $\mathbb{Z}_2$ & Multiplets & $SU(3)_C$ & $SU(2)_L$ & $U(1)_Y$ & $U(1)_X$ & $\mathbb{Z}_2$ \\ \hline 
$l_{1L}=\begin{pmatrix}
\nu_{1L}\\
e_{1L}\end{pmatrix}$ & $\bf 1$ & $\bf 2$ & $-1/2$ & $3z$ & $+$ & $l_{xL}=\begin{pmatrix}
\nu_{xL}\\
e_{xL}\end{pmatrix}$ & $\bf 1$ & $\bf 2$ & $-1/2$ & $-3z$ & $+$ \\ 
$e_{1R}$ & $\bf 1$ & $\bf 1$ & $-1$ & $3z$ & $+$ & $e_{xR}$ & $\bf 1$ & $\bf 1$ & $-1$ & $-3z$ & $+$\\
$q_{\al L}=\begin{pmatrix}
u_{\al L}\\
d_{\al L}\end{pmatrix}$ & $\bf 3$ & $\bf 2$ & $1/6$ & $z$ & $+$ & $q_{3L}=\begin{pmatrix}
u_{3L}\\
d_{3L}\end{pmatrix}$ & $\bf 3$ & $\bf 2$ & $1/6$ & $-z$ & $+$\\
$u_{\al R}$ & $\bf 3$ & $\bf 1$ & $2/3$ & $z$ & $+$ & $u_{3 R}$ & $\bf 3$ & $\bf 1$ & $2/3$ & $-z$ & $+$ \\
$d_{\al R}$ & $\bf 3$ & $\bf 1$ & $-1/3$ & $z$ & $+$ & $d_{3 R}$ & $\bf 3$ & $\bf 1$ & $-1/3$ & $-z$ & $+$\\
$\ph=\begin{pmatrix}
\ph^+\\
\ph^0\end{pmatrix}$ & $\bf 1$ & $\bf 2$  & $1/2$ & $0$ & $+$ &  No data& $\star$ & $\star$ & $\star$ & $\star$ & $\star$ \\
$\nu_R$ & $\bf 1$ & $\bf 1$ & $0$ & $-3z$ & $+$ & $N_R$ & $\bf 1$ & $\bf 1$ & $0$ & $0$ & $-$ \\
$\chi_1$ & $\bf 1$ & $\bf 1$ & $0$ & $2z$ & $+$ & $\chi_2$ & $\bf 1$ & $\bf 1$ & $0$ & $6z$ & $+$\\
$\eta=\begin{pmatrix}
\eta^0\\
\eta^-\end{pmatrix}$ & $\bf 1$ & $\bf 2$ & $-1/2$ & $3z$ & $-$ & $\rho=\begin{pmatrix}
\rho^0\\
\rho^-\end{pmatrix}$ & $\bf 1$ & $\bf 2$ & $-1/2$ & $-3z$ & $-$ \\
\hline\hline
\end{tabular}
\caption[]{\label{tab2}Particle content of proposed model, $\al=1,2$, $x=2,3$, and $z$ is arbitrarily nonzero.}
\ec
\end{table}

We note to the reader that the arguments for the existence of only three fermion generations we present here are similar to those in our recent work \cite{VanLoi:2023kgl} but different from those in the 3-3-1 model \cite{Singer:1980sw,Valle:1983dk,Pisano:1992bxx,Frampton:1992wt,Foot:1994ym,Dong:2006mg,Dong:2006cn,Dong:2007ee,VanDong:2023lbn} and our previous works \cite{Nam:2020twn,VanDong:2022cin,VanLoi:2023utt,VanLoi:2023pkt}, which involve the QCD asymptotic freedom condition. Specifically, the solution we are considering in Eq. (\ref{neuR}) is entirely different from that in our recent work whose three right-handed neutrinos possess $X_{\nu_{aR}}=(3z,-3z,-3z)$ \cite{VanLoi:2023kgl}, as well as the conventional $U(1)_{B-L}$ extension in which three right-handed neutrinos have $(B-L)_{\nu_{aR}}=(-1,-1,-1)$ or $(B-L)_{\nu_{aR}}=(-4,-4,5)$ \cite{Montero:2007cd,VanDong:2023thb}, for $a=1,2,3$.

\subsection{Gauge symmetry breaking and matter parity}
To break the gauge symmetry, we introduce two scalar singlets $\chi_{1,2}$ and a scalar doublet $\ph$ under $SU(2)_L$. The singlets $\chi_{1,2}$ break $U(1)_X$ down to a residual symmetry, labeled $R$, and the doublet $\ph$ that is identical to the SM scalar doublet breaks $SU(2)_L\otimes U(1)_Y$ down to the electromagnetic symmetry $U(1)_Q$ as usual. We would like to emphasize that the $\chi_1$ is necessarily presented to generate the mixing between the first two quark generations and third quark generation through nonrenormalizable operators like $\bar{q}_{\al L}\ph\chi_1 d_{3R}$, for recovering Cabibbo–Kobayashi–Maskawa (CKM) matrix, while $\chi_2$ is included to provide Majorana mass for the right-handed neutrino $\nu_R$ through a coupling $\nu_R\nu_R\chi_2$, responsible for tree-level neutrino mass generation. The scalar multiples and their quantum numbers are also listed in Table \ref{tab2}. Additionally, these scalar multiples develop vacuum expectation values (VEVs), 
\be \langle\chi_1\rangle = \fr{\La_1}{\sqrt2}, \hs \langle\chi_2\rangle=\fr{\La_2}{\sqrt2}, \hs \langle \ph\rangle = \begin{pmatrix}0\\ \fr{v}{\sqrt2}\end{pmatrix},\label{VEVs}\ee
satisfying $\La_{1,2}\gg v$ and $v=246$ GeV for consistency with the SM. 

As a normal transformation, we write the residual symmetry of $U(1)_X$ as $R=e^{i\de X}$ with $\de$ to be a transforming parameter. Because $R$ conserves the vacuums of $\chi_{1,2}$, we have $R\langle\chi_{1,2}\rangle=\langle\chi_{1,2}\rangle$, implying $e^{i\de 2z}=1$ and $e^{i\de 6z}=1$ or $\de=\kappa\pi/z$ for $\kappa$ integer, thus $R=e^{i\kappa\pi X/z}=(-1)^{\kappa X/z}$. From the fifth and eleventh columns of Table \ref{tab2}, it is easy to see that if $\kappa=0$, then $R=1$ for all fields and every $z$, which is the identity transformation. Additionally, the minimal value of $|\kappa|$ that is nonzero and still satisfies $R=1$ for all fields is $2$. Hence, the residual symmetry $R$ is automorphic to a discrete group, such as $\mathcal{Z}_2=\{1,u\}$ with $u=(-1)^{X/z}$ and $u^2=1$. Further, since the spin parity $p_s=(-1)^{2s}$ is always conserved by the Lorentz symmetry, we conveniently multiply the discrete group $\mathcal{Z}_2$ with the spin parity group $S=\{1,p_s\}$ to obtain a new group $\mathcal{Z}_2\otimes S$, which has an invariant discrete subgroup to be 
\be \mathbb{Z}_2=\{1,p\} \ee
with $p=u\times p_s=(-1)^{X/z+2s}$ and $p^2=1$, and thus $\mathcal{Z}_2\otimes S\cong [(\mathcal{Z}_2\otimes S)/\mathbb{Z}_2]\otimes\mathbb{Z}_2$. Because $[(\mathcal{Z}_2\otimes S)/\mathbb{Z}_2]=\{\{1,p\},\{u,p_s\}\}$ is conserved if $\mathbb{Z}_2$ is conserved, we hereafter consider $\mathbb{Z}_2$ to be a residual symmetry instead of $\mathcal{Z}_2$, for convenience. Under $\mathbb{Z}_2$, all the SM fields, $\nu_R$, and $\chi_{1,2}$ are even ($p=1$), whereas $N_R$ is odd ($p=-1$), as presented in the sixth and last columns of Table \ref{tab2}. 

The right-handed neutrino $\nu_R$, through the type-I seesaw mechanism, yields a rank $1$ mass matrix (see below), which makes only a nonzero neutrino mass inappropriate with the experiment \cite{ParticleDataGroup:2022pth}. Therefore, we introduce two additional scalar doublets, $\eta$ and $\rho$, for which $\eta$ couples $l_{1L}$ to $N_R$ while $\rho$ couples $l_{xL}$ to $N_R$, and both are odd under $\mathbb{Z}_2$. This induces appropriate neutrino masses through the scotogenic mechanism \cite{Ma:2006km}, as described by loop-level diagrams in Fig. \ref{fig1}. The quantum numbers of $\eta$ and $\rho$ are displayed in the last two rows of Table \ref{tab2}, respectively. Of course, $\eta$ and $\rho$ have vanished VEVs, preserved by the matter parity.

\subsection{Gauge and scalar sectors}
The gauge bosons acquire masses via the kinetic terms of scalar fields, $\sum_S(D^\mu S)^\dagger (D_\mu S)$ with $S=\phi, \chi_{1,2}$ when the gauge symmetry breaking occurs. Here, the covariant derivative is defined as 
$D_\mu = \pa_\mu + i g_s t_p G_{p\mu} + i g T_j A_{j\mu}+ i g_Y Y B_\mu + i g_X X C_\mu$, in which $(g_s,g,g_Y, g_X)$, $(t_p, T_j, Y, X)$, and $(G_{p\mu}, A_{j\mu}, B_\mu, C_\mu)$ denote coupling constants, generators, and gauge bosons of $(SU(3)_C, SU(2)_L, U(1)_Y, U(1)_X)$ groups, respectively. Because the SM scalar doublet $\ph$ is not charged under $U(1)_X$, while the new scalar singlets $\chi_{1,2}$ do not transform under $SU(2)_L\otimes U(1)_Y$, there is no tree-level mixing between the SM $Z$ boson and the new gauge boson $Z'$.\footnote{A kinetic mixing effect between the $U(1)_{Y, X}$ gauge fields, which is defined by $-\frac{1}{2} \epsilon B_{\mu\nu}C^{\mu\nu}$, has omitted due to the smallness of the mixing parameter $\epsilon$. Indeed, since the SM fermions and additional scalar doublets are charged under the $U(1)_{Y, X}$ groups, the kinetic mixing effect can be generated at one loop level at low energy. The parameter $\epsilon$ in this case is defined by $\epsilon=\frac{g_Yg_X}{24\pi^2}\left[\sum_f(Y_{f_L}+Y_{f_R})X_f\ln\frac{m_r}{m_f}+2\sum_sY_sX_s\ln\frac{m_r}{m_s}\right]$, where $m_r$ is a renormalization scale and $f$ runs over every fermion of the SM with mass $m_f$, while $s=\eta,\rho$ \cite{Bauer:2022nwt}. Then, we easily estimate $|\epsilon|\sim 10^{-3}\times\left(\frac{|z|g_X}{0.1}\right)\ln\left[\left(\frac{m_r}{10^{16}\text{ GeV}}\right)\left(\frac{10^2\text{ GeV}}{m_f}\right)\right]\sim 10^{-3}$.} Hence, we obtain the SM gauge bosons $W^\pm, A, Z$ and the new gauge boson $Z'$ with their masses as
\bea W^\pm &=& \frac{1}{\sqrt2}(A_1\mp iA_2),\hs m^2_W = \fr{g^2v^2}{4} ,\\
 A &=& s_W A_3 + c_W B,\hs m_A=0,\\
Z &=& c_W A_3 - s_W B,\hs m^2_Z = \frac{g^2v^2}{4c^2_W},\\
Z' &=& C,\hs m^2_{Z'}= 4g^2_Xz^2(\La^2_1+9\La^2_2), \eea
where the Weinberg's angle is defined by $\tan(\theta_W) = g_Y/g$, as usual. Additionally, we have labeled $t_W\equiv \tan(\theta_W)$, $s_W\equiv \sin(\theta_W)$, $c_W\equiv \cos(\theta_W)$ for short.

The scalar content includes normal fields ($\ph,\chi_{1,2}$) that induce gauge symmetry breaking and dark fields ($\eta,\rho$), so the scalar potential can be decomposed into two parts, such as $V=V(\ph,\chi_{1,2})+V(\eta,\rho,\text{mix})$, in which
\bea V(\ph,\chi_{1,2}) &=& \mu_1^2 \phi^\dagger \phi + \mu_2^2 \chi_1^* \chi_1 + \mu_3^2 \chi_2^* \chi_2 + \la_1(\phi^\dagger \phi)^2 + \la_2 (\chi_1^* \chi_1)^2 + \la_3 (\chi_2^* \chi_2)^2\crn
&&+(\phi^\dagger \phi)(\la_4\chi_1^* \chi_1 + \la_5\chi_2^* \chi_2)+ \la_6 (\chi_1^* \chi_1)(\chi_2^* \chi_2)+ (\la\chi_1^3\chi_2^*+\mathrm{H.c.}),\\
V(\eta,\rho,\text{mix}) &=& \mu^2_4 \eta^\dagger \eta + \mu^2_5 \rho^\dagger \rho+\la_7(\eta^\dagger \eta)^2+\la_8(\rho^\dagger \rho)^2+\la_9(\eta^\dagger \eta)(\rho^\dagger \rho)+\la_{10}(\eta^\dagger \rho)(\rho^\dagger \eta)\crn
&& +(\ph^\dagger \ph)(\la_{11} \eta^\dagger \eta +\la_{12} \rho^\dagger \rho)+\la_{13} (\eta^\dagger\ph)(\ph^\dagger \eta)+\la_{14} (\rho^\dagger\ph)(\ph^\dagger \rho)\crn
&&(\chi_1^* \chi_1)(\la_{15}\eta^\dagger \eta+\la_{16}\rho^\dagger \rho)+(\chi_2^* \chi_2)(\la_{17}\eta^\dagger \eta+\la_{18}\rho^\dagger \rho)\crn
&&+[\la_{19}(\ph\eta)(\ph\rho)+\mu (\eta^\dagger \rho) \chi_2 + \mathrm{H.c.}],
\eea
where the couplings $\la$'s are dimensionless, whereas $\mu$'s have a mass dimension. Additionally, the hermiticity of the potential requires that these parameters are real, except for $\la,\la_{19}$, and $\mu$, which can be complex. However, without loss of generality we can consider $\la,\la_{19}$, and $\mu$ to be real for the following computation, since their complex phases can be eliminated by redefining the relevant scalar fields, such as $\chi_1\to\chi_1 e^{-i\al_1/3}, \eta\to\eta e^{i(\al_3-\al_2)/2}$, and $\rho\to\rho e^{-i(\al_3+\al_2)/2}$ for $\la=|\la| e^{i\al_1}, \la_{19}=|\la_{19}| e^{i\al_2}$, and $\mu=|\mu| e^{i\al_3}$, which leaves the potential invariant. Note that the VEVs of $\phi$ and $\chi_{1,2}$ may in principle be complex. In such case, since the potential must be invariant under the gauge symmetry, we can do two global rotations $U(1)_Y$ and $U(1)_X$ to eliminate two complex phases associated with the VEVs of $\phi$ and $\chi_2$. Regarding the $\al_0$ complex phase related to the VEV of $\chi_1$, i.e. $\langle\chi_1\rangle=\frac{\La_1}{\sqrt2}e^{i\al_0}$, the minimization of the scalar potential requires $\al_0=k\pi/3$ for $k$ integer, which either conserves spontaneous CP for $k=3p=0,\pm 3, \pm 6,\cdots,$ or violates spontaneous CP for other cases of $k$, i.e. $k=3p\pm 1=\pm 1,\pm 2,\pm 4,\pm 5$, where $p$ is integer. For simplicity, we will consider the VEV of $\chi_1$ conserving CP, say $\langle\chi_1\rangle=\frac{\La_1}{\sqrt2}$ as the outset in Eq. (\ref{VEVs}). The necessary conditions for this scalar potential to be bounded from below and yielding a desirable vacuum structure are $\mu_{1,2,3}^2<0$, $|\mu_1|\ll |\mu_{2,3}|$, $\mu_{4,5}^2>0$, $\la_{1,2,3,7,8}>0$, and others for the scalar self-couplings, which are derived from $V>0$ when two or more than two of scalar fields simultaneously tending to infinity.

Expanding the neutral scalar fields around their VEVs as $\ph^0=(v+S+iA)/\sqrt2$, $\chi_{1,2}=(\La_{1,2}+S_{1,2}+iA_{1,2})/\sqrt2$, $\eta^0=(R_\eta+iI_\eta)/\sqrt2$, and $\rho^0=(R_\rho+iI_\rho)/\sqrt2$, and then substituting them into the scalar potential, we get the potential minimum conditions,
\bea
2\la_1v^2+\la_4\La_1^2+\la_5\La_2^2+2\mu_1^2 &=& 0,\\
2\la_2\La_1^2+\la_4v^2+\la_6\La_2^2+3\la\La_1\La_2+2\mu_2^2 &=& 0,\\
\la\La_1^3+(2\la_3 \La_2^2+\la_5v^2+\la_6\La_1^2+2\mu_3^2)\La_2 &=& 0.
\eea
Hence, we obtain a mass-squared matrix of $CP$-even normal scalars ($S,S_{1,2}$) as
\be M_S^2=\left(\begin{array}{ccc} 2\la_1v^2 & \la_4v\La_1 & \la_5v\La_2 \\
\la_4v\La_1 & \fr 1 2 (4\la_2\La_1+3\la\La_2)\La_1 & \fr 1 2 (2\la_6\La_1\La_2+3\la\La_1^2) \\
\la_5v\La_2 & \fr 1 2 (2\la_6\La_1\La_2+3\la\La_1^2) & \fr{1}{2\La_2}(4\la_3\La_2^3-\la\La_1^3)\end{array}\right). \ee
Since $v\ll\La_{1,2}$, the elements in the first row and column of $M_S^2$ are significantly smaller than those in the rest. This allows us to use the seesaw approximation to diagonalize $M_S^2$ and separate the light state $S$ from the heavy states $S_{1,2}$. Taking a new basis as ($H,\mathcal{H}_1,\mathcal{H}_2$) for which $H$ is decoupled as a physical field, we get
\be H\simeq S-\ep_1 S_1-\ep_2 S_2 \ee
with its mass to be
\be m_H^2\simeq 2\la_1v^2-(\ep_1 \la_4\La_1+\ep_2 \la_5\La_2)v. \ee
Here, the mixing parameters are given by
\bea \ep_1 &=& \frac{[\la(\la_4\La_1^3+3\la_5\La_1\La_2^2)-2(2\la_3\la_4-\la_5\la_6)\La_2^3]v}{2[3\la^2\La_1^3\La_2+\la(\la_2\La_1^4-3\la_3\La_2^4+3\la_6\La_1^2\La_2^2)-(4\la_2\la_3-\la_6^2)\La_1\La_2^3]},\\
\ep_2 &=& \frac{[3\la(\la_4\La_1^2-\la_5\La_2^2)-2(2\la_2\la_5-\la_4\la_6)\La_1\La_2]v\La_2}{2[3\la^2\La_1^3\La_2+\la(\la_2\La_1^4-3\la_3\La_2^4+3\la_6\La_1^2\La_2^2)-(4\la_2\la_3-\la_6^2)\La_1\La_2^3]}, \eea
which are small as suppressed by $v/\La_{1,2}$. The remaining states $\mathcal{H}_1\simeq \ep_1 S + S_1$ and $\mathcal{H}_2\simeq \ep_2 S + S_2$ mix by themselves via a $2\times 2$ submatrix. Diagonalizing this submatrix, we get two physical fields,
 \be H_1 = c_\xi \mathcal{H}_1 - s_\xi \mathcal{H}_2, \hs H_2 = s_\xi \mathcal{H}_1 + c_\xi \mathcal{H}_2,\ee
 with corresponding masses,
 \bea
m^2_{H_{1,2}} &=& \frac{1}{4\La_2}\left\{4\la_3\La_2^3-\la\La_1^3+(4\la_2\La_1+3\la\La_2)\La_1\La_2\right.\crn
&&\left.\mp\sqrt{[4\la_3\La_2^3-\la\La_1^3-(4\la_2\La_1+3\la\La_2)\La_1\La_2]^2+4(2\la_6\La_2+3\la\La_1)^2\La_1^2\La_2^2}\right\}. \eea
The mixing angle $\xi$ is given by 
\be t_{2\xi} = \fr{2(2\la_6\La_2+3\la\La_1)\La_1\La_2}{4\la_3\La_2^3-\la\La_1^3-(4\la_2\La_1+3\la\La_2)\La_1\La_2}. \ee
The mass of Higgs boson $H$ is in weak scale like the SM Higgs boson, so $H$ is identified with the SM Higgs boson, whereas $H_{1,2}$ are the new Higgs bosons, heavy in the $\La_{1,2}$ scale. We note that the presence of the mixing parameters $\ep_{1,2}$ not only results in deviations of the couplings of the SM Higgs boson to the SM fermions and gauge bosons from the ones predicted by the SM but also opens windows for new-physics search. Hence, these parameters are generally constrained by the measurements of the discovered Higgs production cross section, its decay branching ratio, and the null results in current searches at the LHC \cite{ParticleDataGroup:2022pth}. Imposing an individual bound $\ep_{1,2}\lesssim 0.2$ \cite{Buttazzo:2018qqp,CidVidal:2018eel,CMS:2018amk}, we obtain $\La_{1,2}\gtrsim 1.23$ TeV, given that the relevant scalar couplings are of the same order of magnitude.

For the $CP$-odd normal scalars $A$ and $A_{1,2}$, we directly obtain a massless eigenstate, $G_Z=A$, which is identical to the Goldstone boson eaten by the SM $Z$ boson. On the other hand, the scalars $A_{1,2}$ mix by themselves via a $2\times 2$ matrix. Diagonalizing this matrix, we get two relate fields,
\be G_{Z'} =\frac{\La_1A_1+3\La_2A_2}{\sqrt{\La_1^2+9\La_2^2}},\hs \mathcal{A}= \frac{3\La_2A_1-\La_1A_2}{\sqrt{\La_1^2+9\La_2^2}}, \ee
in which $G_{Z'}$ is a Goldstone boson eaten by the new neutral gauge boson $Z'$, whereas $\mathcal{A}$ is a physical pseudoscalar with a heavy mass at the $\La_{1,2}$ scale,  
\be m^2_\mathcal{A}= -\frac{\la(\La_1^2+9\La_2^2)\La_1}{2\La_2}.\ee
The requirement of positive squared mass implies the parameter $\la$ to be negative.

For the dark fields $R_{\eta,\rho}$ and $I_{\eta,\rho}$, they mix in each pair, such as
\bea V &\supset& \fr 1 2 \begin{pmatrix} R_\eta & R_\rho \end{pmatrix}
\begin{pmatrix} M^2_\eta & \fr{\mu \La_2}{\sqrt{2}}+\fr{\la_{19}v^2}{2}\\
\fr{\mu \La_2}{\sqrt{2}}+\fr{\la_{19}v^2}{2} & M^2_\rho \end{pmatrix}
\begin{pmatrix} R_\eta \\ R_\rho\end{pmatrix}\crn
&& +\fr 1 2  \begin{pmatrix} I_\eta & I_\rho \end{pmatrix}
\begin{pmatrix} M^2_\eta & \fr{\mu \La_2}{\sqrt{2}}-\fr{\la_{19}v^2}{2}\\
\fr{\mu \La_2}{\sqrt{2}}-\fr{\la_{19}v^2}{2} & M^2_\rho \end{pmatrix}
\begin{pmatrix} I_\eta \\ I_\rho\end{pmatrix}, \eea 
where $M^2_\eta=\mu^2_4+\fr{\la_{11}}{2}v^2+\fr{\la_{15}}{2}\La_1^2+\fr{\la_{17}}{2}\La_2^2$ and $M^2_\rho=\mu^2_5+\fr{\la_{12}}{2}v^2+\fr{\la_{16}}{2}\La_1^2+\fr{\la_{18}}{2}\La_2^2$. Defining two mixing angles $\theta_{R,I}$ as \be t_{2R,2I}=\fr{\sqrt{2}\mu\La_2\pm\la_{19}v^2}{M^2_{\rho}-M^2_\eta},\label{dsmix}\ee 
we obtain four physical fields,
\bea R_1 &=& c_R R_\eta -s_R R_\rho,\hs R_2=s_R R_\eta +c_R R_\rho,\\
I_1 &=& c_I I_\eta -s_I I_\rho,\hs I_2=s_I I_\eta +c_I I_\rho, \eea 
and their masses,
\bea && m^2_{R_{1}}\simeq M^2_\eta + \fr{(\sqrt{2}\mu\La_2+\la_{19} v^2)^2}{4(M^2_\eta - M^2_\rho)},\hs m^2_{R_{2}}\simeq M^2_\rho - \fr{(\sqrt{2}\mu\La_2+\la_{19} v^2)^2}{4(M^2_\eta - M^2_\rho)},\label{remass}\\
&& m^2_{I_{1}}\simeq M^2_\eta + \fr{(\sqrt{2}\mu\La_2-\la_{19} v^2)^2}{4(M^2_\eta - M^2_\rho)},\hs m^2_{I_{2}}\simeq M^2_\rho - \fr{(\sqrt{2}\mu\La_2-\la_{19} v^2)^2}{4(M^2_\eta - M^2_\rho)},\label{immass}\eea 
given that $\mu\La_2\sim \la_{19}v^2\ll M^2_{\eta,\rho}\sim \bar{M}^2$ and $\bar{M}\sim\mathcal{O}(1)$ TeV. 

Concerning the charged scalars $\ph^\pm$, $\eta^\pm$, and $\rho^\pm$, we directly obtain a massless eigenstate, $G_W^\pm \equiv \ph^\pm$, which is identical to the Goldstone boson eaten by the SM $W$ boson, while $\eta^\pm$ and $\rho^\pm$ mix by themselves via a $2\times 2$ matrix. From here, we obtain two charged physics scalars with their masses to be heavy at the $\La_{1,2}$ scale, such as
\bea H_1^\pm &=& c_\theta \eta^\pm -s_\theta \rho^\pm,\hs m^2_{H_1^\pm}\simeq M^2_\eta+\frac{\la_{13}}{2}v^2,\label{chargedmass}\\
H_2^\pm &=& s_\theta \eta^\pm +c_\theta \rho^\pm,\hs m^2_{H_2^\pm}\simeq M^2_\rho+\frac{\la_{14}}{2}v^2, \eea
assuming $\mu\ll\La_{1,2}$. The mixing angle $\theta$ is small, given by
\be t_{2\theta}\simeq\fr{\sqrt{2}\mu\La_2}{M^2_\rho-M^2_\eta}.\ee

Last, but not least, we comment that the mass splitting between the neutral and charged scalar components of the two scalar doublets $\eta$ and $\rho$ contributes to the oblique parameter $T$ at one-loop level, thus the $\rho_0$ parameter, for the electroweak precision tests, i.e., $\Delta\rho\equiv \rho_0-1\simeq\al T\simeq \fr{1}{16\pi^2v^2}[\mathcal{F}(m^2_{H_1^\pm},m^2_{R_1})+\mathcal{F}(m^2_{H_2^\pm},m^2_{R_2})]$, where the loop function is defined as $\mathcal{F}(x^2,y^2)=x^2+y^2-\frac{4x^2y^2}{x^2-y^2}\ln\frac{x}{y}$ \cite{Grimus:2007if,Haber:2010bw}. It is checked that the contribution agrees with the $1\sigma$ range of the global fit $\rho_0=1.00031\pm 0.00019$ \cite{ParticleDataGroup:2022pth} if the mass splitting is around 10 to 60 GeV. 

\subsection{Fermion mass}
When the scalar multiplets develop VEVs, fermion masses and mixing between different fermion generations are generated through Yukawa interactions. For charged fermions,  they are given by
\bea\mathcal{L}&\supset&h^d_{\al\bet}\bar{q}_{\al L}\ph d_{\bet R}+h^d_{33}\bar{q}_{3L}\ph d_{3R}+\frac{h^d_{\al 3}}{\La_c}\bar{q}_{\al L}\ph\chi_1 d_{3R}+\frac{h^d_{3\bet}}{\La_c}\bar{q}_{3L}\ph\chi_1^* d_{\bet R}\crn
&&+h^u_{\al\bet}\bar{q}_{\al L}\tilde{\ph} u_{\bet R}+h^u_{33}\bar{q}_{3L}\tilde{\ph} u_{3R}+\frac{h^u_{\al 3}}{\La_c}\bar{q}_{\al L}\tilde{\ph}\chi_1 u_{3R}+\frac{h^u_{3\bet}}{\La_c}\bar{q}_{3L}\tilde{\ph}\chi_1^* u_{\bet R}\crn
&&+h^e_{11}\bar{l}_{1L}\ph e_{1R}+h^e_{xy}\bar{l}_{xL}\ph e_{yR}+\frac{h^e_{1y}}{\La_c}\bar{l}_{1L}\ph\chi_2 e_{yR}+\frac{h^e_{x1}}{\La_c}\bar{l}_{xL}\ph\chi_2^* e_{1R}+\mathrm{H.c.},\label{yuk}
\eea
where $h$'s is dimensionless, $\tilde{\ph}=i\sigma_2\ph^*$ with $\sigma_2$ to be the second Pauli matrix, and $\La_c$ denotes a NP (or cutoff) scale that defines the effective interactions. Additionally, the mixing between the first two quark generations and third quark generation arises only from nonrenormalizable operators. This induces the CKM elements $V_{cb}$ and $V_{ub}$ to be naturally small in agreement with the experiment \cite{ParticleDataGroup:2022pth}. Note that in this sector there are uniquely six dimension-5 operators that are invariant under the gauge symmetry, as shown above. From the above interactions, we obtain mass matrices for down-type quarks, up-type quarks, and charged leptons, 
\bea 
[M_q]_{\al\beta} &=& -h^q_{\al\beta}\frac{v}{\sqrt2}, \hs [M_q]_{33}=-h^q_{33}\frac{v}{\sqrt2}, \hs \left[M_q\right]_{\al 3} = -h^q_{\al 3}\fr{v\La_1}{2\La_c}, \hs [M_q]_{3\beta}= -h^q_{3\beta}\frac{v\La_1}{2\La_c},\\
\left[M_e\right]_{11} &=& -h^e_{11}\frac{v}{\sqrt2}, \hs [M_e]_{xy} = -h^e_{xy}\frac{v}{\sqrt2},\hs \left[M_e\right]_{1y} = -h^e_{1y}\frac{v\La_2}{2\La_c}, \hs [M_e]_{x1} = -h^e_{x1}\frac{v\La_2}{2\La_c},  \eea
where $q=u,d$. Diagonalizing these mass matrices by the bi-unitary transformations, one by one, we get the mass of the relative particles, such as
\bea \mathrm{diag}(m_d,m_s,m_b) &=& V_{d_L}^\dag M_d V_{d_R},\\
\mathrm{diag}(m_u,m_c,m_t) &=& V_{u_L}^\dag M_u V_{u_R},\\
\mathrm{diag}(m_e,m_\mu,m_\tau) &=& V_{e_L}^\dag M_e V_{e_R},\eea
in which $V_{d_{L,R}}$, $V_{u_{L,R}}$, and $V_{e_{L,R}}$ are unitary matrices, respectively linking the physical states, $d'=(d,s,b)^T$, $u'=(u,c,t)^T$, and $e'=(e,\mu,\tau)^T$, to the respective gauge states, $d=(d_1,d_2,d_3)^T$, $u=(u_1,u_2,u_3)^T$, and $e=(e_1,e_2,e_3)^T$, namely
\be d_{L,R}=V_{d_{L,R}}d'_{L,R}, \hs u_{L,R}=V_{u_{L,R}}u'_{L,R}, \hs e_{L,R}=V_{e_{L,R}}e'_{L,R}. \ee
The CKM matrix is then given by $V_{\mathrm{CKM}}=V_{u_L}^\dag V_{d_L}$.

\begin{figure}[h]
\centering
\includegraphics[scale=1]{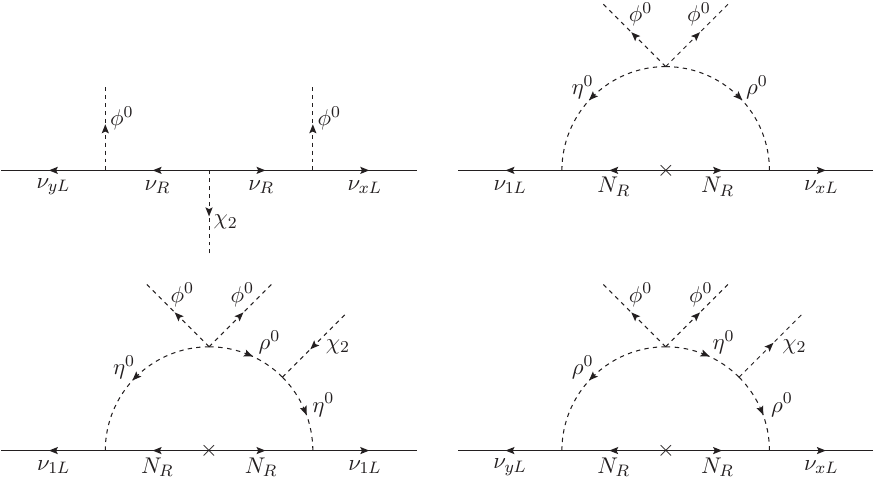}
\caption[]{\label{fig1}Scoto-seesaw neutrino mass generation governed by matter parity, where $x,y=2,3$ are generation indices.}
\end{figure}

For the neutrinos, their Yukawa interactions are given by
\be
\mathcal{L} \supset h_x\bar{l}_{xL}\tilde{\ph}\nu_R+\fr 1 2 f \bar{\nu}_R^c\nu_R\chi_2+k_1\bar{l}_{1L}\eta N_R+k_x\bar{l}_{xL}\rho N_R-\fr 1 2 M_N\bar{N}_R^c N_R+\mathrm{H.c.},
\ee
where $h_x$, $f$, and $k_{1,x}$ are the coupling constants and $M_N$ is the Majorana mass of the fermion $N_R$. From here, we obtain a mass Lagrangian at the tree level as
\be \mathcal{L}\supset -\fr 1 2 \begin{pmatrix} \bar{\nu}_{aL} & \bar{\nu}^c_R\end{pmatrix}\begin{pmatrix} 0 & M_D\\
M_D^T & M_\nu \end{pmatrix}\begin{pmatrix} \nu^c_{bL} \\
\nu_R\end{pmatrix} + \mathrm{H.c.},\label{seesaw}\ee
where $a,b=1,2,3$ are generic generation indexes, and $M_D=-\frac{v}{\sqrt2}(0, h_2, h_3)^T$ is a Dirac mass matrix, while $M_\nu=-f\frac{\La_2}{\sqrt2}$ is a Majorana mass. Because of $\La_2\gg v$, i.e., $M_\nu\gg M_D$, the mass matrix in Eq. (\ref{seesaw}) can be diagonalized by using the seesaw approximation to separate the light states ($\nu_{aL}$) from the heavy state ($\nu_R$), such as  
\be \mathcal{L}\supset -\fr 1 2 \bar{\nu}_{aL}(m_\nu^{\mathrm{tree}})_{ab}\nu^c_{bL}-\fr 1 2 M_\nu \bar{\nu}^c_R\nu_R + \mathrm{H.c.},\ee 
in which the seesaw-induced neutrino mass matrix is given by
\be (m_\nu^{\mathrm{tree}})_{ab}= -\fr{(M_D)_{a1} (M_D^T)_{1b}}{M_\nu}\label{seesawmass},\ee
which corresponds to the tree-level diagram (left-upper) in Fig. \ref{fig1}. Notice that $m_\nu^{\mathrm{tree}}$ is a neutrino mass matrix of rank 1, yielding only one massive light neutrino.

In addition to the seesaw contribution, the light neutrino masses in our model receive a scotogenic contribution from the loop-level diagrams (right-upper and lower) in Fig. 1 with the dark fields $\eta$, $\rho$, and $N_R$. In mass base, these loop diagrams are determined by the following Lagrangian,
\bea \mathcal{L} &\supset& \fr{k_1}{\sqrt{2}}\bar{\nu}_{1L}(c_R R_1+s_R R_2+i c_I I_1+i s_I I_2)N_R\crn
&&+\fr{k_x}{\sqrt{2}}\bar{\nu}_{xL}(-s_R R_1+c_R R_2-i s_I I_1+i c_I I_2)N_R-\fr 1 2 M_N \bar{N}_R^cN_R + \mathrm{H.c.}\crn
&&- \fr 1 2 m^2_{R_1}R_1^2 - \fr 1 2 m^2_{R_2}R_2^2 - \fr 1 2 m^2_{I_1}I_1^2 - \fr 1 2 m^2_{I_2}I_2^2.\eea
Therefore, the loop-induced neutrino mass matrix can be written as
\be (m_\nu^{\mathrm{rad}})_{ab}= \mathcal{F}_{\eta\eta}M_N (K)_{a1}(K^T)_{1b}+\mathcal{F}_{\rho\rho}M_N (\mathcal{K})_{a1}(\mathcal{K}^T)_{1b}+\mathcal{F}_{\eta\rho}M_N [(K)_{a1}(\mathcal{K}^T)_{1b}+(\mathcal{K})_{a1}(K^T)_{1b}], \ee
where $K=(k_1,0,0)^T$, $\mathcal{K}=(0,k_2,k_3)^T$, and the loop factors $\mathcal{F}_{\eta\eta,\rho\rho,\eta\rho}$ are 
\bea \mathcal{F}_{\eta\eta}&=&\fr{1}{32\pi^2}\left(\fr{c^2_R m^2_{R_1} \ln \fr{M^2_N}{m^2_{R_1}}}{M^2_N-m^2_{R_1}}- \fr{c^2_I m^2_{I_1} \ln \fr{M^2_N}{m^2_{I_1}}}{M^2_N-m^2_{I_1}}+ \fr{s^2_R m^2_{R_2} \ln \fr{M^2_N}{m^2_{R_2}}}{M^2_N-m^2_{R_2}}- \fr{s^2_I m^2_{I_2} \ln \fr{M^2_N}{m^2_{I_2}}}{M^2_N-m^2_{I_2}}\right),\\
\mathcal{F}_{\rho\rho}&=&\fr{1}{32\pi^2}\left(\fr{s^2_R m^2_{R_1} \ln \fr{M^2_N}{m^2_{R_1}}}{M^2_N-m^2_{R_1}}- \fr{s^2_I m^2_{I_1} \ln \fr{M^2_N}{m^2_{I_1}}}{M^2_N-m^2_{I_1}}+ \fr{c^2_R m^2_{R_2} \ln \fr{M^2_N}{m^2_{R_2}}}{M^2_N-m^2_{R_2}}- \fr{c^2_I m^2_{I_2} \ln \fr{M^2_N}{m^2_{I_2}}}{M^2_N-m^2_{I_2}}\right),\\
\mathcal{F}_{\eta\rho}&=&\fr{1}{64\pi^2}\left[\left(\fr{m^2_{R_2} \ln \fr{M^2_N}{m^2_{R_2}}}{M^2_N-m^2_{R_2}}-\fr{m^2_{R_1} \ln \fr{M^2_N}{m^2_{R_1}}}{M^2_N-m^2_{R_1}}\right)s_{2R}+\left(\fr{m^2_{I_1} \ln \fr{M^2_N}{m^2_{I_1}}}{M^2_N-m^2_{I_1}}- \fr{m^2_{I_2} \ln \fr{M^2_N}{m^2_{I_2}}}{M^2_N-m^2_{I_2}}\right)s_{2I}\right]. \eea
In summary, the total light neutrino mass matrix involving both type-I seesaw and scotogenic contributions is given by 
\be (m_\nu^{\mathrm{tot}})_{ab}=(m_\nu^{\mathrm{tree}})_{ab}+(m_\nu^{\mathrm{rad}})_{ab}, \ee
and thus, its mass eigenvalues can be defined as
\be \mathrm{diag}(m_1,m_2,m_3)= V^T_{\nu_L}m_\nu^{\mathrm{tot}}V_{\nu_L}, \ee
where $V_{\nu_L}$ is a unitary matrix, connecting the physical neutrino states $\nu'_L=(\nu_{eL},\nu_{\mu L},\nu_{\tau L})^T$ to the gauge neutrino states $\nu_L=(\nu_{1L},\nu_{2L},\nu_{3L})^T$ as $\nu_L=V_{\nu_L}\nu'_L$. The Pontecorvo-Maki-Nakagawa-Sakata (PMNS) matrix is then given by $V_\mathrm{PMNS}=V_{\nu_L}^\dag V_{e_L}$. It is stressed that $m_\nu^{\mathrm{tot}}$ is a neutrino mass matrix of rank 3, yielding three massive light neutrinos appropriate to experiment \cite{ParticleDataGroup:2022pth}. Additionally, in a simplified framework, i.e., $h_2=h_3$ and $k_1=k_2=k_3$, $m_\nu^{\mathrm{tot}}$ is a rank 2 mass matrix, generating only two massive light neutrinos, which are still sufficient to explain the observed neutrino masses \cite{ParticleDataGroup:2022pth}.

Based on the results obtained in Eqs. (\ref{dsmix}), (\ref{remass}), and (\ref{immass}), it is clear that both the mixing angles-squared $\theta^2_{R,I}$ and the dark scalar mass splittings, $|m^2_{R_\al}-m^2_{I_\al}|/m^2_{R_\al,I_\al}$ with $\al=1,2$, are proportional to $\la^2_{19}v^4/M_{\eta,\rho}^4\ll 1$, assuming $\mu\La_2\sim \la_{19}v^2$ and $M_{\eta,\rho}\sim \bar{M}\sim\mathcal{O}(1)\text{ TeV}$. This gives us an estimate for the loop-induced neutrino masses, such as $[m_\nu^{\mathrm{rad}}]_{11,xy}\sim v^4(k\la_{19})^2M_N/(32\pi^2\bar{M}^4)\sim 0.1\times (M_N/\text{TeV})(k\la_{19}/10^{-4})^2$ eV, provided that $\bar{M}\sim\mathcal{O}(1)$ TeV and $k\sim k_a$ for $a=1,2,3$. Taking the experimental value $m_\nu\sim 0.1$ eV, the above estimate implies $M_N\sim 1$ TeV and $k\la_{19}\sim 10^{-4}$. Additionally, $[m_\nu^{\mathrm{rad}}]_{1y,x1}\sim [m_\nu^{\mathrm{rad}}]_{11,xy} \times s_{R,I}$, so they are small too. On the other hand, the seesaw-induced neutrino mass in Eq. (\ref{seesawmass}) is proportional to $h^2v^2/f\La_2$, given that $h\sim h_x$ for $x=2,3$, so the experimental value $m_\nu\sim 0.1$ eV indicates $h\sim 10^{-5.9}$ that is close to the Yukawa coupling constant of the electron if $f\sim 0.1$ and $\La_2\sim 10$ TeV.

\subsection{Fermion-gauge boson interaction}
The interaction of gauge bosons with fermions arises from fermion kinetic terms, $\sum_F \bar{F}i\gamma^\mu D_\mu F$, where $F$ runs over fermion multiplets. The covariant derivative is defined, in terms of physical fields, as $D_\mu = \pa_\mu + i g_s t_p G_{p\mu} + i g s_W Q A_\mu+ i g(T_+W^+_\mu+\mathrm{H.c.})+\frac{ig}{c_W}(T_3-s^2_WQ)Z_\mu+ i g_X X Z'_\mu$, where $T_\pm=(T_1\pm iT_2)/\sqrt2$ is weight-raising/lowering operator, respectively. Hence, the gluons, the photon, and the SM $Z$-like boson interact with normal fermions as in the SM. The charged currents of the quarks are also not modified, but that of the leptons is now given by
\be \mathcal{L}\supset -\frac{g}{\sqrt2}\bar{\nu}_{iL}\gamma^\mu [V_{\mathrm{PMNS}}]_{ij}e_{jL} W^+_\mu+\mathrm{H.c.} \ee
Here and in further investigation, we use indices $i,j=1,2,3$ to label physical states, such as $\nu_i=\nu_e,\nu_\mu,\nu_\tau$, $e_i=e,\mu,\tau$, $u_i=u,c,t$, and $d_i=d,s,b$ for $i=1,2,3$.

Because the $X$-charge is not universal for every flavor of neutrinos, charged leptons, up-type quarks, and down-type quarks, in contrast to the usual charges as $Q$, $T_3$, and $Y$, our model predicts flavor-changing processes at tree level in both the quark and lepton sectors, associated with the new gauge boson $Z'$, in addition to flavor-conserving processes. Using the unitary condition, $V_{u_{L,R}}^\dag V_{u_{L,R}}=V_{d_{L,R}}^\dag V_{d_{L,R}}=V_{e_{L,R}}^\dag V_{e_{L,R}}=V_{\nu_L}^\dag V_{\nu_L}=1$, we get the relative interactions,
\bea \mathcal{L}&\supset& -zg_X(\bar{u}_i\gamma^\mu u_i+\bar{d}_i\gamma^\mu d_i-3\bar{e}_i\gamma^\mu e_i-3\bar{\nu}_{iL}\gamma^\mu \nu_{iL}-3\bar{\nu}_R\gamma^\mu \nu_R)Z'_\mu \crn
&&+ \Gamma^{\nu_L}_{ij}\bar{\nu}_{iL}\gamma^\mu \nu_{jL}Z'_\mu+[(\Gamma^{u_L}_{ij}\bar{u}_{iL}\gamma^\mu u_{jL}+\Gamma^{d_L}_{ij}\bar{d}_{iL}\gamma^\mu d_{jL}+ \Gamma^{e_L}_{ij}\bar{e}_{iL}\gamma^\mu e_{jL})Z'_\mu+(L\to R)],\label{Z'inter}\eea
where $i,j$ are summed, and we have defined
\be \Gamma^{q_L}_{ij}= 2zg_X[V^*_{q_L}]_{3i}[V_{q_L}]_{3j},\hs \Gamma^{\ell_L}_{ij} = -6zg_X[V^*_{\ell_L}]_{1i}[V_{\ell_L}]_{1j}, \label{fvio}\ee
for $q=u$ or $d$ and $\ell=e$ or $\nu$. The terms of the first line in Eq. (\ref{Z'inter}) describe the flavor-conserving interactions, while the remaining terms give rise to the flavor-changing interactions for $i\neq j$.

\section{\label{collider}Collider bounds}
The new neutral gauge boson $Z'$ predicted by our model is heavy at the TeV scale and couples to both quarks and leptons. In this section, we will focus on the flavor-conserving interactions of $Z'$ with fermions and investigate the potential for discovering $Z'$ at two significant experiments: the large electron-positron (LEP) collider \cite{Carena:2004xs,ALEPH:2005ab,ALEPH:2013dgf} and the large hadron collider (LHC) \cite{ATLAS:2019erb,ATLAS:2019fgd,CMS:2021ctt}. 

\subsection{LEP}
At the LEP, when the mass of the $Z'$ boson is larger than the largest collider energy (approximately 209 GeV for LEP-II), it is not directly generated in electron-positron collisions. However, it can still be detected indirectly through the processes $e^+e^-\to\bar{f}f$ mediated by $Z'$, where $f$ are various SM fermions, by observing the deviations from the relative predictions of the SM. For convenience, we parametrize such processes by effective four-fermion contact interactions, such as
\be \mathcal{L}_{\text{eff}}=\frac{1}{1+\delta_{ef}}\frac{1}{m^2_{Z'}}\sum_{A,B=L,R}C_{e_A}^{Z'}C_{f_B}^{Z'}(\bar{e}\gamma_\mu P_Ae)(\bar{f}\gamma^\mu P_Bf), \ee
where $\delta_{ef}=1(0)$ for $f=e(\neq e)$, $P_{L,R}=\fr 1 2 (1\mp\ga_5)$, and $C_{f_{L,R}}^{Z'}$ are chiral gauge couplings of $Z'$ with $f$. Notice that all the SM fermions are vector-like under $U(1)_X$. From the interactions in Eq. (\ref{Z'inter}), we extract the relative couplings, namely
\be C^{Z'}_e=-C^{Z'}_{\mu,\tau}=C^{Z'}_{\nu_e}=-C^{Z'}_{\nu_\mu,\nu_\tau} = 3C^{Z'}_{u,c,d,s}=-3C^{Z'}_{t,b}=-3zg_X.\label{Zpcouplings}\ee 

All lower limits of the scale of these contact interactions, labeled $\La^\pm_{ff}$, have been reported by LEP-II, in which $\La^+_{ff}$ for $C_{e_A}^{Z'}C_{f_B}^{Z'}>0$ and $\La^-_{ff}$ for $C_{e_A}^{Z'}C_{f_B}^{Z'}<0$ \cite{Carena:2004xs}. The strongest constraint for our model, where $\mu$ and $\tau$ have the same $X$ charges and $e$ has the $X$ charge of opposite sign, comes from the channel $e^+e^-\to\mu^+\mu^-(\tau^+\tau^-)$ with $\La^-_{\mu\mu}\geq 16.3$ TeV \cite{ALEPH:2013dgf}. Hence, we obtain a relevant constraint as 
\be \fr{m_{Z'}}{|z|g_X}\gtrsim 13.79 \text{ TeV}.\label{LEPcon} \ee
Furthermore, using projected limits on $\La_{\mu\mu}$ reported in Ref. \cite{LCCPhysicsWorkingGroup:2019fvj}, we estimate prospective constraints on $m_{Z'}/|z|g_X$ at the International Linear Collider (ILC) with center of mass energies $\sqrt{s}=250\text{ GeV}, 500\text{ GeV}$, and $1\text{ TeV}$ to be $99.02$ TeV, $167.56$ TeV, and $280.97$ TeV, respectively. 

\subsection{LHC}
At the LHC, the $Z'$ boson can be directly generated in hadron colliders through the channel $\bar{q}q\to Z'$ and subsequently decayed to quark pairs (dijet) or lepton pairs (dilepton), where the most significant decay channel is $Z'\to\bar{l}l$ with $l=e,\mu$ because of well-understood backgrounds \cite{ATLAS:2019erb,CMS:2021ctt} and that it signifies a boson $Z'$ having both couplings to quarks and leptons like our model. Using the narrow width approximation, the cross section for the relevant process is given by \cite{Accomando:2010fz}
\be
\sigma(pp\rightarrow Z'\rightarrow \bar{l}l) \simeq \fr 1 3 \sum_q\fr{dL_{\bar{q}q}}{dm^2_{Z'}}\hat{\sigma}(\bar{q}q\rightarrow Z')\text{BR}(Z'\rightarrow \bar{l}l),\label{sigppll}\ee
where the parton luminosities are written as $dL_{\bar{q}q}/dm^2_{Z'}$, which can be extracted from Ref. \cite{Martin:2009iq}, and $\hat{\sigma}(\bar{q}q\rightarrow Z')$ is the peak cross-section approximated as $\hat{\sigma}(\bar{q}q\rightarrow Z')\simeq (C^{Z'}_q)^2\pi/3$. The branching ratio of $Z'$ decaying into the lepton pair is given by $\text{BR}(Z'\rightarrow\bar{l}l)= \Gamma(Z'\rightarrow\bar{l}l)/\Gamma_{Z'}$ with $\Gamma(Z'\rightarrow\bar{l}l)\simeq m_{Z'}(C_l^{Z'})^2/12\pi$ and 
\be
\Gamma_{Z'}\simeq\frac{m_{Z'}}{12\pi}\sum_f N_C(f)(C_f^{Z'})^2+\frac{m_{Z'}}{8\pi}(C^{Z'}_{\nu_e})^2,
\ee
assuming that the decay channels of $Z'$ into right-handed neutrinos and new scalars negligibly contribute to the total width of $Z'$. Above, $f$ denotes the SM charged fermions, $N_f$ is the color number of the fermion $f$, $\Theta$ is the step function, and the relevant couplings are given in Eq. (\ref{Zpcouplings}).

\begin{figure}[h]
\centering
\includegraphics[scale=0.41]{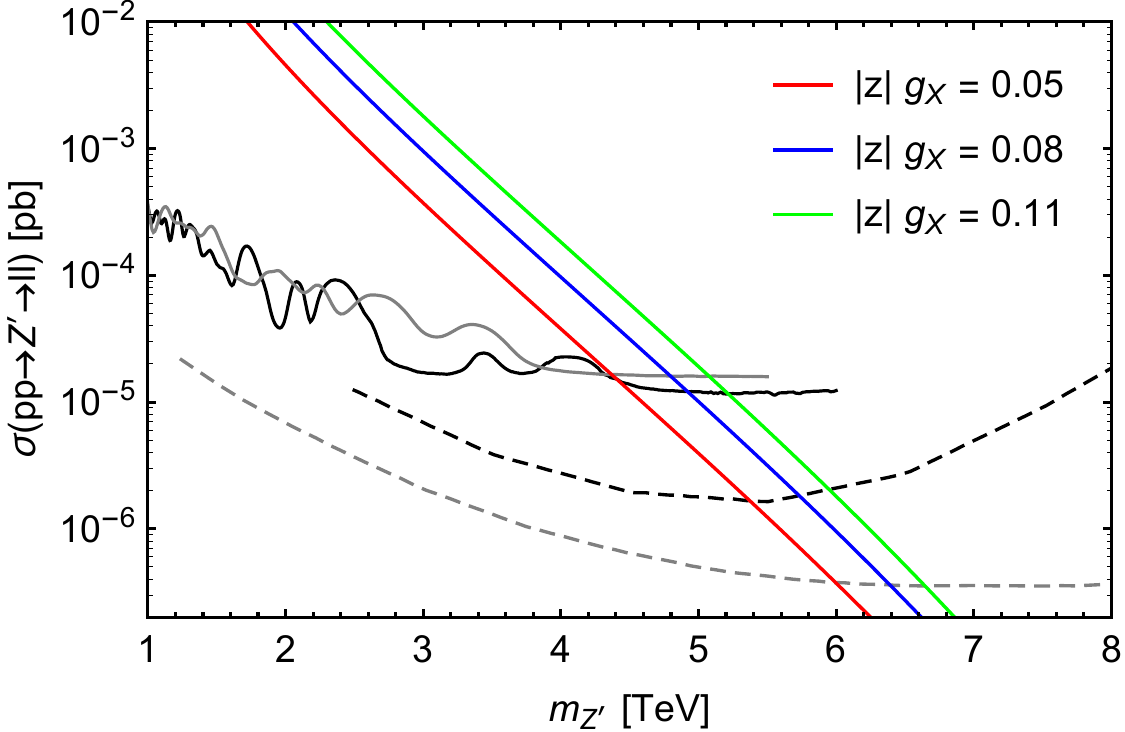}
\includegraphics[scale=0.41]{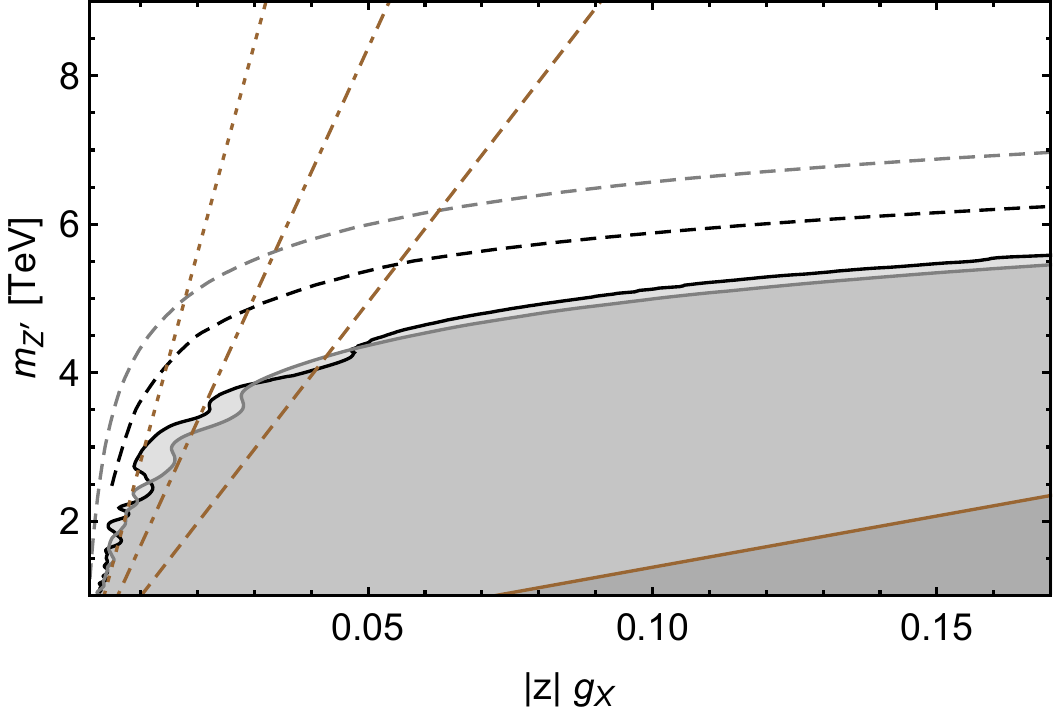}
 \caption[]{\label{fig2}Left panel: dilepton production cross-section as a function of $Z'$-boson mass for various values of the $|z|g_X$ product. The black (gray) curve shows the upper bound on the cross section obtained from ATLAS (CMS) \cite{ATLAS:2019erb,CMS:2021ctt}. Right panel: the black (gray, brown) curve denotes the lower bound on the $Z'$-boson mass obtained from ATLAS (CMS, LEP-II). The parameter space according to shaded regions is excluded. Additionally, prospective sensitivities are shown in dashed black (ATLAS) \cite{ATLAS:2018tvr}, dashed gray (CMS) \cite{CMS:2022gho}, dashed brown (ILC250), dot-dashed brown (ILC500), and dotted brown (ILC1000) \cite{LCCPhysicsWorkingGroup:2019fvj}.}
\end{figure}

In the left panel of Fig. \ref{fig2}, we plot the cross section for process $\sigma(pp\rightarrow Z'\rightarrow \bar{l}l)$ as a function of $Z'$-boson mass according to three distinct values of the product $|z|g_X$. Additionally, we include the upper limits on the cross section of this process as a black (gray) curve, which is observed by the ATLAS-2019 for $\Gamma/m=3\%$ \citep{ATLAS:2019erb} (the CMS-2019 for $\Gamma/m=0.6\%$ \citep{CMS:2021ctt}). It is clear that the lower bounds on the $Z'$-boson mass are $4.41, 4.93$, and $5.21$ TeV corresponding to $|z|g_X = 0.05, 0.08$, and $0.11$. Further, in the right panel of this figure, we show the lower bound of $Z'$-boson mass defined by the ATLAS (CMS) for a range of $|z|g_X$ as the black (gray) curve. We also add the lower bound on the $Z'$-boson mass obtained from the LEP-II (brown curve) for comparison. The available regions for the $Z'$-boson mass lie above these curves. It is easy to see that the constraints from the ATLAS and CMS are much stronger than one from the LEP-II.

Last but not least, to probe the $Z'$ boson at future hadron-hadron colliders, one can extrapolate the current collider searches to a future high-luminosity and/or high-energy collider by adopting the method described in Refs. \cite{Papucci:2014rja,Thamm:2015zwa}. Additionally, search prospects for heavy bosons in fermionic final states at high luminosity LHC are also presented by the ATLAS and CMS collaborations. In the panels of Fig. \ref{fig2}, we show prospective sensitivities predicted by the ATLAS (dashed black) and CMS (dashed gray) at 14 TeV center-of-mass energy and a total luminosity of 3 ab$^{-1}$ \cite{ATLAS:2018tvr, CMS:2022gho}. For completeness, we also describe prospective sensitivities predicted by the ILC with $\sqrt{s}=250\text{ GeV}, 500\text{ GeV}, 1\text{ TeV}$ (dashed, dot-dashed, dotted brown lines) to the right panel. We see significantly higher mass reaches in comparison to the current searches.

\section{\label{flavor}Flavor anomalies}
As an assumption, we align the quark mixing to the down quark sector and the lepton mixing to the neutral lepton sector, i.e., $V_{d_L}=V_\mathrm{CKM}$ and $V_{\nu_L}^\dag=V_\mathrm{PMNS}$, so $V_{u_L}=V_{e_L}=1$. Additionally, we assume $V_{u_R,d_R,e_R}\simeq V_{u_L,d_L,e_L}$ since the current experiment does not define these right-handed fermion mixing matrices and the gauge symmetry under consideration that contains the baryon and lepton numbers may obey a left-right symmetry at high energy. Such a choice of basis implies that the Yukawa coupling matrices $h^u_{ab}$ and $h^e_{ab}$ associated with the last two lines of Eq. (\ref{yuk}) are diagonal, i.e., $h^u_{ab}=-\fr{\sqrt2}{v}\text{diag}(m_u,m_c,m_t)$ and $h^e_{ab}=-\fr{\sqrt2}{v}\text{diag}(m_e,m_\mu,m_\tau)$.  

We would like to note that the CKM matrix can be parameterized by three mixing angles $\theta_{12,13,23}$ and the $CP$-violating phase $\delta$ \cite{Chau:1984fp}. Further, these mixing angles can be defined via the Wolfenstein parameters $\la,A,\bar{\rho},\bar{\eta}$ \cite{Wolfenstein:1983yz,Buras:1994ec,Charles:2004jd}, i.e., 
\bea \sin\theta_{12}=\la,\hs \sin\theta_{23}= A\la^2, \hs \sin\theta_{13}=A\la^3\sqrt{\bar{\rho}^2+\bar{\eta}^2}/(1-\la^2/2).
\eea  
The values of the Wolfenstein parameters and known input parameters associated with quark flavor phenomenology, which will be used in our numerical study, are listed in Table \ref{input-par}.

\begin{table}[h]
	\begin{centering}
	\begin{tabular}{|c|c|c|c|}
		\hline Input parameters & Values  & Input parameters & Values \tabularnewline	\hline 
	$\la$ & $0.22519(83) $  \cite{UTfit:2022hsi} & $A$ & $0.828(11)$ \cite{UTfit:2022hsi} \tabularnewline
	$\bar{\rho}$ & $0.1609(95)$ \cite{UTfit:2022hsi} & $\bar{\eta}$ & $0.347(10)$ \cite{UTfit:2022hsi} \tabularnewline
	$ m_{u}$ & $2.16(7) \ \text{MeV}$  \cite{ParticleDataGroup:2022pth} &  $m_{d}$ & $4.70(7)  \ \text{MeV} $ \cite{ParticleDataGroup:2022pth}\tabularnewline
	$m_{c} $ & $1.2730(46)  \ \text{GeV} $ \cite{ParticleDataGroup:2022pth} & $ m_{s}$ & $93.5(8) \ \text{MeV}$  \cite{ParticleDataGroup:2022pth} \tabularnewline
	$m_{t}$ & $172.57(29) \  \text{GeV} $ \cite{ParticleDataGroup:2022pth} & $m_{b}$ & $4.183(7)  \ \text{GeV} $ \cite{UTfit:2022hsi} \tabularnewline
	$f_K$ & $155.7(3) \ \text{MeV}$ \cite{FlavourLatticeAveragingGroupFLAG:2021npn} &  $ m_K$ & $497.611(13) \ \text{MeV}$  \cite{ParticleDataGroup:2022pth} \tabularnewline
	$f_{B_d}$ & $190.0(1.3) \ \text{MeV}$ \cite{FlavourLatticeAveragingGroupFLAG:2021npn} &  $ m_{B_d}$ & $5279.72(8) \ \text{MeV}$  \cite{ParticleDataGroup:2022pth} \tabularnewline
	$f_{B_s}$ & $230.3(1.3) \ \text{MeV}$ \cite{FlavourLatticeAveragingGroupFLAG:2021npn} &  $ m_{B_s}$ & $5366.93(10) \ \text{MeV}$  \cite{ParticleDataGroup:2022pth} \tabularnewline
	$N(E_{\ga})$ & $3.3\times 10^{-3} $ \cite{Misiak:2020vlo}& $C_7^{\text{SM}}(\mu_b=2 \ \text{GeV})$ & $-0.3636  $\cite{Misiak:2006ab,Czakon:2006ss,Misiak:2020vlo} \tabularnewline\hline 
	\end{tabular}\par
	\protect\caption{\label{input-par}Numerical values of known input parameters for quark flavors.}
	\end{centering}
	\end{table}

\subsection{\label{QFV}Meson oscillations and quark transitions}
\begin{figure}[h]
\centering
\begin{subfigure}[b]{0.9\textwidth}
\centering
\includegraphics[scale=1.4]{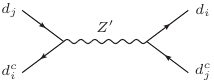}
\caption[]{Tree-level diagrams for $C_{K,B_d,B_s}$.}
\label{fig3a}
\end{subfigure}

\begin{subfigure}[b]{0.9\textwidth}
\centering
\includegraphics[scale=1.4]{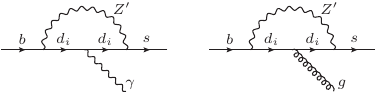}
\caption[]{One-loop diagrams for $C_7$ (left) and $C_8$ (right).}
\label{fig3b}
\end{subfigure}

\begin{subfigure}[b]{0.9\textwidth}
\centering
\includegraphics[scale=1.4]{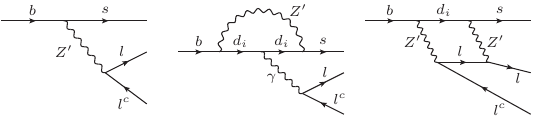}
\caption[]{Tree-level, penguin, and box diagrams for $C_9$.}
\label{fig3c}
\end{subfigure}

\caption[]{Diagrams for quark flavor-changing processes induced by $Z'$, where $d_{i(j)}=d,s,b$ for $i(j)=1,2,3$ and $l=e,\mu$.}
\label{fig3}
\end{figure}

Since the quark generations are not universal under the additional gauge group $U(1)_X$, the present model predicts flavor changing processes in the quark sector associated with the additional gauge boson $Z'$, such as the neutral meson oscillations $K^0$--$\bar{K}^0$ and $B_{d,s}^0$--$\bar{B}_{d,s}^0$ at the tree level, and the quark transition $b\to s$ at both tree and loop levels, as descried by Feynman diagrams in Fig. \ref{fig3}. The effective Hamiltonian relevant for these processes can be written as \cite{Buchalla:1995vs}
\be \mathcal{H}^{\text{quark}}_{\text{eff}} = \sum_{X=K,B_d,B_s}C_X^2(\mathcal{O}_X^2+\mathcal{O}_X^{'2}+2\mathcal{O}_X\mathcal{O}^{'}_X)-\fr{4G_F}{\sqrt{2}}V^{*}_{ts}V_{tb}\sum_{Y=7,8,9}C_Y(\mathcal{O}_Y+\mathcal{O}_Y^{'}),\label{quarkHamil}\ee
where $G_F$ is the Fermi constant, $V_{ts,tb}$ are the CKM matrix elements, and the primed operators $\mathcal{O}^{'}_{X,Y}$ are chirally flipped counterpart of unprimed operators $\mathcal{O}_{X,Y}$, i.e., $P_L\leftrightarrow P_R$. 

The operators and Wilson coefficients in the first summation in Eq. (\ref{quarkHamil}) are given by
\bea 
\mathcal{O}^{(')}_K &=&\bar{d}\ga^{\mu}P_{L(R)}s,  \hs \mathcal{O}^{(')}_{B_d}=\bar{d}\ga^{\mu}P_{L(R)}b, \hs \mathcal{O}^{(')}_{B_s} =\bar{s}\ga^{\mu}P_{L(R)}b ,\\
C_K &=& \Gamma^{d_L}_{12}/m_{Z'}, \hs C_{B_d} = \Gamma^{d_L}_{13}/m_{Z'}, \hs C_{B_s} = \Gamma^{d_L}_{23}/m_{Z'}.  
\eea 
Hence, the contribution of NP to the neutral meson mass differences is estimated as \cite{Gabbiani:1996hi,Langacker:2000ju}
\bea \Delta m_K^\text{NP} &=& \fr{2}{3} \mathrm{Re}[C_K^2]\left[\fr 1 2 - \left(\fr{m_K}{m_d+m_s}\right)^2\right]m_K f^2_K,\\
\Delta m_{B_d}^\text{NP} &=&\Delta m_K^{Z'}({C_K\to C_{B_d}},m_K\to m_{B_d},m_s\to m_b,f_K\to f_{B_d}),\\
\Delta m_{B_s}^\text{NP} &=&\Delta m_K^{Z'}({C_K\to C_{B_s}},m_K\to m_{B_s},m_d\to m_b,f_K\to f_{B_s}). \eea
The measurement results and SM predictions for the meson mass differences are respectively label as $\Delta m^\text{Exp}_{B_{d,s},K}$ and $\Delta m^\text{SM}_{B_{d,s},K}$, and their current values are presented in Table \ref{quarkflavor}. For the $B^0_{d,s}$--$\bar{B}^0_{d,s}$ meson systems, at $1\sigma$ range, we have $\Delta m_{B_d}^\text{SM}/\Delta m_{B_d}^\text{Exp}=1.0712(1\pm 0.0535)$ and $\Delta m_{B_s}^\text{SM}/\Delta m_{B_s}^\text{Exp}=1.0566(1\pm 0.0458)$. Imposing $\Delta m_{B_d,B_s}^\text{Exp}=\Delta m_{B_d,B_s}^\text{SM}+\Delta m_{B_d,B_s}^\text{NP}$, we obtain the following constraints,
\be \fr{\Delta m_{B_d}^\text{NP}}{\Delta m_{B_d}^\text{Exp}}\in [-0.1286,-0.0139], \hs \fr{\Delta m_{B_s}^\text{NP}}{\Delta m_{B_s}^\text{Exp}}\in [-0.1050,-0.0082]. \label{BdBs}\ee
With the $K^0$--$\bar{K}^0$ meson system, since the lattice QCD calculations for long-distance effect are not well controlled, we require that the present theory contributes about $30\%$ to $\Delta m_K$, i.e., $\Delta m_K^\text{SM}/\Delta m_K^\text{Exp}=1(1\pm 0.3)$, which then translates to a constraint as
\be \fr{\Delta m_K^\text{NP}}{\Delta m_K^\text{Exp}}\in [-0.3,0.3], \ee
taking $\Delta m_K^\text{Exp}=\Delta m_K^\text{SM}+\Delta m_K^\text{NP}$.

\begin{table}[h!]
	\begin{centering}
		\begin{tabular}{|c|c|c|}
			\hline
			Observation & SM prediction & Experimental value  
\tabularnewline \hline 
$\Delta m_K$ & $0.467\times 10^{-2}\text{ ps}^{-1}$ \cite{Buras:2016dxz} & $0.5293(9)\times 10^{-2}\text{ ps}^{-1}$ \cite{ParticleDataGroup:2022pth}
\tabularnewline \hline 
$\Delta m_{B_d}$ & $0.543(29)\text{ ps}^{-1}$ \cite{Lenz:2019lvd} & $0.5069(19)\text{ ps}^{-1}$ \cite{ParticleDataGroup:2022pth}
\tabularnewline \hline
$\Delta m_{B_s}$ & $18.77(86)\text{ ps}^{-1}$ \cite{Lenz:2019lvd} & $17.765(6)\text{ ps}^{-1}$ \cite{ParticleDataGroup:2022pth}
\tabularnewline \hline
$\text{BR}(\bar{B}\to X_s\gamma)$ & $3.40(17)\times 10^{-4}$ \cite{Misiak:2020vlo} & $3.49(19)\times 10^{-4}$ \cite{ParticleDataGroup:2022pth}
\tabularnewline \hline
$R_K$ & $1.00(1)$ \cite{Bordone:2016gaq} & $0.949^{+0.042+0.022}_{-0.041-0.022}$ \cite{LHCb:2022qnv,LHCb:2022vje}
\tabularnewline \hline
$R_{K^*}$ & $1.00(1)$ \cite{Bordone:2016gaq} & $1.027^{+0.072+0.027}_{-0.068-0.026}$ \cite{LHCb:2022qnv,LHCb:2022vje}
\tabularnewline \hline
		\end{tabular}
		\par
			\protect\caption{\label{quarkflavor}The SM predictions and experimental values for quark flavor changing observables.}
	\end{centering}
\end{table}

The second summation in Eq. (\ref{quarkHamil}) relates to the quark transition $b\to s$, for which the relevant operators are defined as
\bea 
\mathcal{O}^{(')}_7 &=&\fr{e}{16\pi^2}m_b(\bar{s}\sigma^{\mu\nu}P_{R(L)} b)F_{\mu\nu}, \\ \mathcal{O}^{(')}_8&=&\fr{g_s}{16\pi^2}m_b(\bar{s}\sigma^{\mu\nu}T^a P_{R(L)} b)G^a_{\mu\nu}, \\ 
\mathcal{O}^{(')}_{9}&=&\fr{e^2}{16\pi^2}(\bar{s}\ga^{\mu}P_{L(R)}b)(\bar{l}\ga_{\mu}l), \eea
with $e$ to be the electromagnetic coupling constant, $e=gs_W$, and $l=e,\mu$. For the Wilson coefficients $C_Y$, we decompose each of them as the sum of four distinct pieces, namely $C_Y=C^\text{tree}_Y+C^\text{loop}_Y+C^\text{penguin}_Y+C^\text{box}_Y$ for which the superscripts indicate the style of relative diagrams. At the scale $\mu_{Z'}=\mathcal{O}(m_{Z'})$, we obtain 
\bea 
C^\text{loop}_{7} (\mu_{Z'}) &\simeq& - \fr{2m_W^2}{9g^2V^{*}_{ts}V_{tb}}\fr{1}{m_{Z'}^2}(\Ga^{*d_L}_{12}\Ga^{d_L}_{13}+C^{Z'}_s\Ga^{d_L}_{23}-2\Ga^{*d_L}_{32}C^{Z'}_b) ,\\ 
C^\text{loop}_{8} (\mu_{Z'}) &\simeq& -3C^\text{loop}_{7}  (\mu_{Z'}), \hs C^\text{tree}_{9,l}(\mu_{Z'})\simeq -\fr{32m_W^2 \pi^2}{e^2g^2V^*_{ts}V_{tb}}\fr{\Ga^{d_L}_{23}C^{Z'}_l}{m_{Z'}^2}, \\ 
C_{9,l}^\text{penguin}(\mu_{Z'}) &\simeq& -\fr{m_W^2}{9g^2 V_{ts}^*V_{tb}}\fr{1}{m_{Z'}^2}\left(\Ga^{*d_L}_{12}\Ga^{d_L}_{13}\ln{\frac{m^4_d}{m^4_{Z'}}}+C^{Z'}_s\Ga^{d_L}_{23}\ln{\frac{m^4_s}{m^4_{Z'}}}+\Ga^{*d_L}_{32}C^{Z'}_b\ln{\frac{m^4_b}{m^4_{Z'}}}\right), \\ 
C^\text{box}_{9,l}(\mu_{Z'}) &\simeq& -\fr{m_W^2}{2e^2g^2V^{*}_{ts}V_{tb}}\fr{1}{m_{Z'}^2}(\Ga^{*d_L}_{12}\Ga^{d_L}_{13}+C^{Z'}_s\Ga^{d_L}_{23}+\Ga^{*d_L}_{32}C^{Z'}_b)(C^{Z'}_l)^2. \eea
Here, we use 't Hooft gauge $\zeta=1$ for calculating the diagrams. With the diagrams in subfigure \ref{fig3b}, we calculate on shell, i.e., $q^2=0$, $p_s^2=m_s^2$, and $p_b^2=m_b^2$. Since $m_s\ll m_b$, we set the $s$ quark mass to be zero, $m_s=0$, and keep the $b$ quark mass at the linear order, i.e., $m_b^2=0$. Additionally, we calculate in the limit $m_{d_i}^2/m_{Z'}^2,m_l^2/m_{Z'}^2\ll 1$ since $m_{d_i,l}\sim \mathcal{O}(1) \ \text{GeV} \ll m_{Z'}\sim \mathcal{O}(1)$ TeV, for simplicity. Notice that under this limit other loop diagrams associated with the Goldstone boson $G_{Z'}$ are suppressed by factors $m_{d_i}^2/m_{Z'}^2,m_l^2/m_{Z'}^2\ll 1$, hence we can safely ignore the box diagrams associated with the Goldstone boson $G_{Z'}$ and keep only the ones associated with the new gauge boson $Z'$. 

In the presence of NP, the $\bar{B}\to X_s\gamma$ branching ratio is given by \cite{Gambino:2001ew,Buras:2011zb}
\be
\text{BR}(\bar{B}\to X_s\ga)=\fr{6\al_{\text{em}}}{\pi C}\left|\fr{V_{ts}^*V_{tb}}{V_{cb}}\right|^2 \left[|C_7(\mu_b)|^2+|C^{'}_7(\mu_b)|^2+N(E_{\gamma})\right]\text{BR}(\bar{B}\rightarrow X_c e\bar{\nu}) , \label{bra1}
\ee
where $\al_\text{em}$ is the electromagnetic fine structure constant, $\al_\text{em}=e^2/4\pi$, and $C$ is the semileptonic phase-space factor, and $N(E_{\ga})$ is a nonperturbative contribution, estimated at the level of around $4\%$ of the branching ratio \cite{Misiak:2020vlo}, and BR$(\bar{B}\rightarrow X_c e\bar{\nu})$ is the branching ratio for semileptonic decay. The coefficients $C_{7}^{(')}(\mu_b)$ are evaluated at the matching scale $\mu_b=2$ GeV by running down from the higher scale $\mu_{Z'}=\mathcal{O}(m_{Z'})$ via the renormalization group equations. These coefficients can be split as
\bea C_{7}(\mu_b)=C_{7}^{\text{SM}}(\mu_b)+C_{7}^{\text{NP}}(\mu_b),\hs  C^{'}_{7}(\mu_b)=C_{7}^{\text{NP}}(\mu_b), \eea 
in which $C_{7}^\text{SM}(\mu_b)$ is the SM Wilson coefficient, calculated up to next-to-next-leading order of QCD corrections, while $C^\text{NP}_{7}(\mu_{b})$ is the NP Wilson coefficient, calculated at leading order \cite{Buras:2011zb} as 
\bea 
C^\text{NP}_{7}(\mu_{b})=\ka_7C^\text{loop}_7(\mu_{Z'})+\ka_8C_8^\text{loop}(\mu_{Z'})+\Delta_{Z'}^\text{current}(\mu_{b}).\label{C7NP}
\eea
The last term in Eq. (\ref{C7NP}) results from the mixing of new neutral current-current operators, generated by the exchange of $Z'$ with the dipole operators $\mathcal{O}_{7,8}$,
\be \Delta_{Z'}^\text{current}(\mu_{b}) = -\frac{2m_W^2}{m_{Z'}^2g^2V^*_{ts}V_{tb}} \left[\sum_{A,f}\ka_{LA}^f \Gamma^{*d_L}_{23}C^{Z'}_f+\sum_A\hat{\ka}^d_{LA} \Gamma^{*d_L}_{21}\Gamma_{31}^{d_A}\right] \ee
for $A=L,R$, and $f=u,c,t,d,s,b$. The coefficients $\ka$ and $\hat{\ka}$ are NP magic numbers and their numerical values are given in Ref. \cite{Buras:2011zb}. Considering the ratio among the experimental and SM values for this branching ratio (see Table \ref{quarkflavor}), we obtain a constraint at $1\sigma$ range as
\be 1+2\frac{|C_7^\text{NP}|^2+C_7^\text{SM}\text{Re}[C_7^\text{NP}]}{|C_7^\text{SM}|^2+N(E_\gamma)}=1.0265(1\pm 0.0739).  \ee

The $R_K$ and $R_{K^*}$ lepton flavor universality testing ratios measured by LHCb Collaboration (see Table \ref{quarkflavor}) are defined in terms of the Wilson coefficients $C_9^l= C^\text{tree}_{9,l}+C_{9,l}^\text{penguin}+C^\text{box}_{9,l}$ in the range of squared dilepton mass $q^2=[1.1,6.0]$ as \cite{Cornella:2021sby}
\bea  \fr{R_K}{R_K^{\text{SM}}} &=& \frac{1+0.48\text{Re}[C_9^\mu]+0.06|C_9^\mu|^2}{1+0.48\text{Re}[C_9^e]+0.06|C_9^e|^2},\\
\fr{R_{K^*}}{R_{K^*}^{\text{SM}}} &=& \frac{1+0.36\text{Re}[C_9^\mu]+0.06|C_9^\mu|^2}{1+0.36\text{Re}[C_9^e]+0.06|C_9^e|^2}, \eea
where $R_{K,K^*}^{\text{SM}}$ are the SM predictions for $R_{K,K^*}$. The numerical values for $R_{K,K^*}$ predicted by SM and measured by experiment are shown in Table \ref{quarkflavor}.

\begin{figure}[h]
\centering
\includegraphics[scale=0.41]{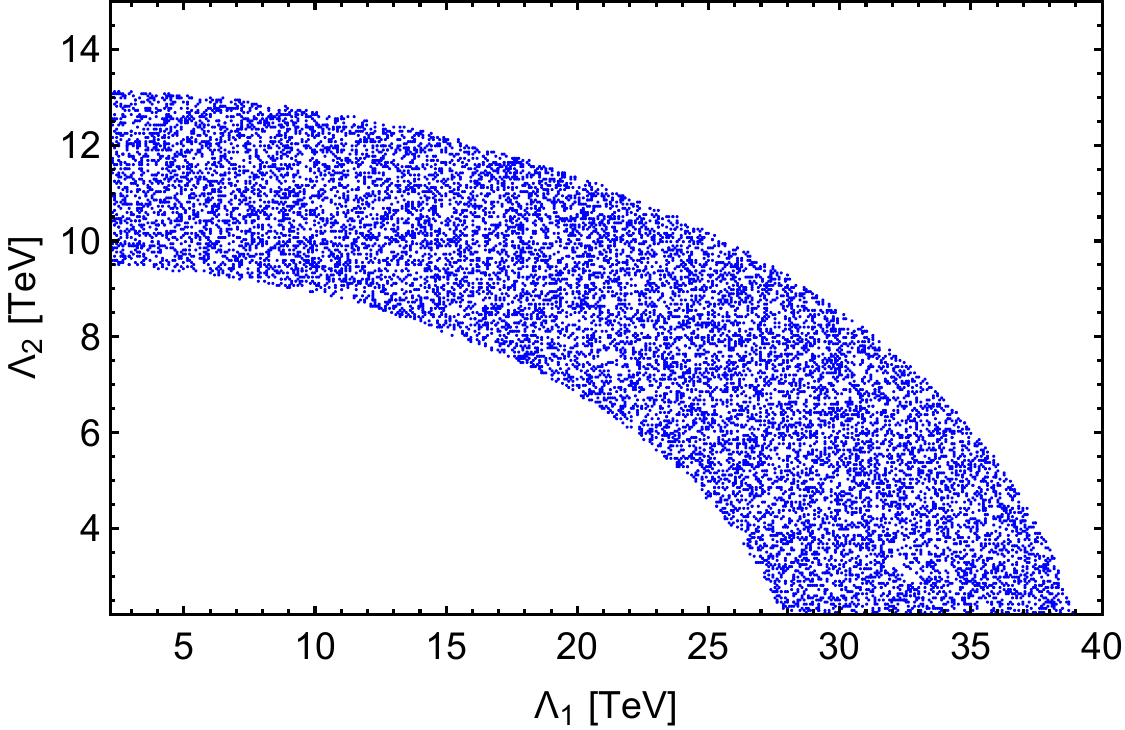}
\includegraphics[scale=0.41]{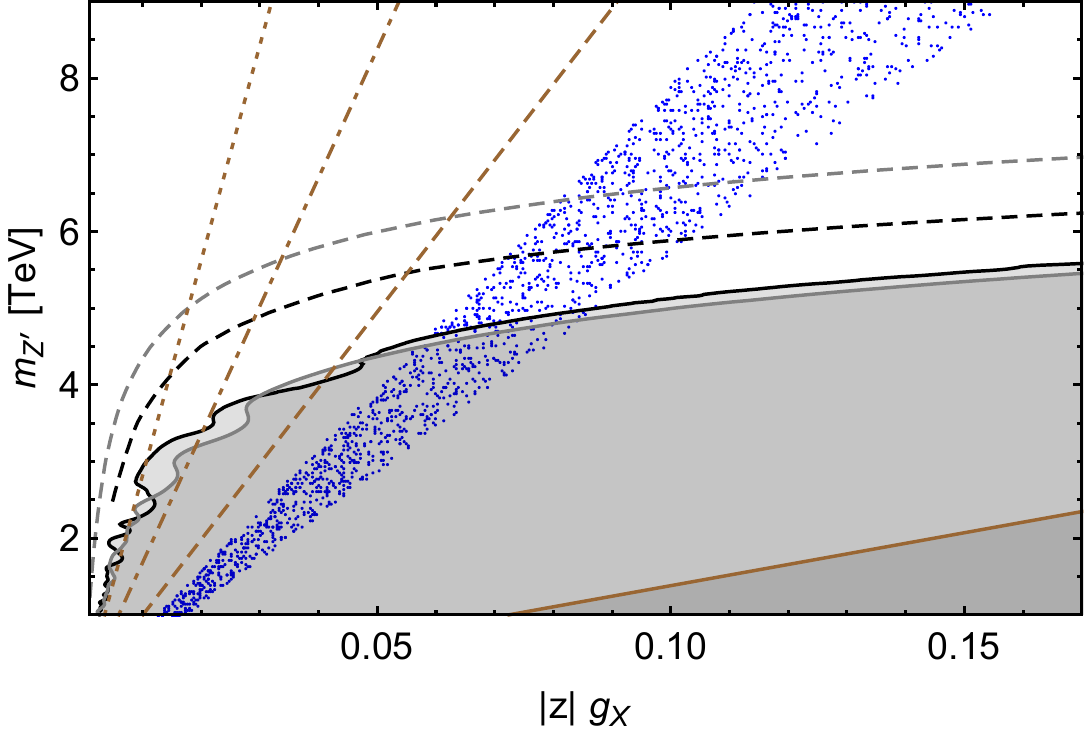}
 \caption[]{\label{fig4}The correlations between the $\La_1$ and $\La_2$ VEVs (left panel), and between the mass of new gauge boson $m_{Z'}$ and the product $|z|g_X$ (right panel). The collider bounds presented in Fig. \ref{fig2} are included to the right panel for completeness.}
\end{figure}

Doing numerical analysis for the observables mentioned above, we use the known input parameters listed in Table \ref{input-par} and scan the free parameters in ranges $|z|g_X\in [10^{-3},1]$ and $\La_{1,2}\in [2.2,50]$ TeV. Here, the scanning ranges ensure the mixing parameters $\ep$ and $\ep_{1,2}$ to be suppressed \cite{ParticleDataGroup:2022pth}. Especially, the lower bound of scanning range of $\La_{1,2}$ results from the LEP constraint defined in Eq. (\ref{LEPcon}). Results are shown in the plane of $\La_1$ versus $\La_2$ (left panel) and $m_{Z'}$ versus $|z|g_X$ (right panel) in Fig. \ref{fig4}, where the blue points satisfy all the constraints. The left panel demonstrates that the viable points are limited in the ranges $2.2\text{ TeV}\leq\La_1\lesssim 39.32\text{ TeV}$ and $2.2\text{ TeV}\leq\La_2\lesssim 13.14\text{ TeV}$. Additionally, the behavior of $\La_1$ and $\La_2$ is inverse, namely if the value of $\La_1$ increases then $\La_2$ decreases and vice versa. We note that the lower and upper bounds of the blue region relate to the lepton flavor universality testing ratio $R_{K^*}$ and the neutral meson mass difference $\Delta m_{B_d}$, respectively. This can be interpreted as follows: if both $\La_{1,2}$ are made small (below the blue region), $m_{Z'}$ is reduced, then the magnitude of $C_9^l$ is enhanced, and so $R_{K^*}$ is smaller than its lower experimental bound. On the other side, if both $\La_{1,2}$ are made large (above the blue region), then $m_{Z'}$ is enhanced, which leads to the decrease of $C_{B_d}$. Therefore, the magnitude of $\Delta m_{B_d}^\text{NP}/\Delta m_{B_d}^\text{Exp}$ is smaller than the 0.0139 value shown in Eq. (\ref{BdBs}). Further, we find a relevant constraint as
\be 57.09\text{ TeV}\lesssim 2\sqrt{\La_1^2+9\La_2^2}\equiv \fr{m_{Z'}}{|z|g_X}\lesssim 78.92 \text{ TeV}.\label{quarkcon}\ee
The lower bound in Eq. (\ref{quarkcon}) is more stringent and quite larger than the one determined in Eq. (\ref{LEPcon}). In the right panel, to obtain a lower bound for $m_{Z'}$, we include the results from the ATLAS, CMS, and LEP-II, which are shown in the right panel of Fig. \ref{fig2}. We see that the viable points are limited in the ranges
\be m_{Z'} \gtrsim 4.64\text{ TeV}, \hs |z|g_X\gtrsim 0.059. \ee

It is noteworthy that due to the flavor-violating couplings in both quark and lepton sectors, the model additionally contains other observables in $b\to s\bar{\nu}\nu$ transitions such as branching ratio of decays $B^+\to K^+\bar{\nu}\nu$ and $B\to K^{*0}\bar{\nu}\nu$ at the tree-level by $Z'$ exchange, besides the SM contribution generated by $W$ boson at loop and box diagrams. The effective Hamiltonian describes these kinds of observables is given as follows
\be
\mathcal{H}_{\text{eff}}^{b\to s\bar{\nu}\nu}=-\fr{4G_F}{\sqrt{2}}V_{ts}^*V_{tb}\left(C^{\text{SM},\nu_i\nu_i}_{LL}\mathcal{O}^{W,\nu_i\nu_i}_{LL}+C^{Z',\nu_i\nu_j}_{L(R),L}\mathcal{O}^{Z',\nu_i\nu_j}_{L(R),L}\right),  \ee 
where $\nu_{i(j)}=\nu_{e,\mu,\tau}$. Additionally, the first operator is given by $\mathcal{O}^{W,\nu_i\nu_i}_{LL}=\fr{e^2}{16\pi^2}(\bar{s}\ga_{\mu}P_{L}b)(\bar{\nu}_{i}\ga^{\mu}P_L\nu_{i})$ with the Wilson coefficient to be the same for different neutrino flavors, $C^{\text{SM},\nu_i\nu_i}_{LL}=-2X_t/s_W^2$, because the lepton flavor universality is preserved in SM. Here, $X_t$ has been performed up to NLO QCD and two-loop electroweak corrections $X_t=1.469(17)$ \cite{Buras:2014fpa}. For the rest, the operator $\mathcal{O}^{Z',\nu_i\nu_j}_{L(R),L}$ and their Wilson coefficients induced by $Z'$ interaction read 
\bea \mathcal{O}^{Z',\nu_i\nu_j}_{L(R),L} &=& \fr{e^2}{16\pi^2}(\bar{s}\ga_{\mu}P_{L(R)}b)(\bar{\nu}_{i}\ga^{\mu}P_L\nu_{j} ), \\
C^{Z',\nu_i\nu_j}_{L(R),L} &=&\begin{cases}
	-\fr{4\pi^2\sqrt{2}\Ga^{d_{L(R)}}_{23}\Ga^{\nu_L}_{ij}}{G_FV_{ts}^*V_{tb}e^2m_{Z'}^2}  &\text{if } i \neq j  \\
	-\fr{4\pi^2\sqrt{2}\Ga^{d_{L(R)}}_{23}C^{Z'}_{\nu_i}}{G_FV_{ts}^*V_{tb}e^2m_{Z'}^2} &\text{if } i=j  \\
\end{cases},
\eea 
in which the couplings $\Ga^{\nu_L}_{ij}$ and $C^{Z'}_{\nu_i}$ have been defined by Eqs. (\ref{fvio}) and (\ref{Zpcouplings}), respectively. The ratio between the model prediction for BR$(B\to K\bar{\nu}\nu)$ and corresponding SM prediction is given as contributions of both diagonal and off-diagonal neutrino flavors \cite{Browder:2021hbl}
\be 
R_{K}^{\nu}=\fr{1}{3}\sum_{\nu_i}\left|1+\fr{C^{Z',\nu_i\nu_i}_{LL}+C^{Z',\nu_i\nu_i}_{RL}}{C^{\text{SM},\nu_i\nu_i}_{LL}}\right|^2 + \sum_{\nu_i\neq\nu_j}\left|\fr{C^{Z',\nu_i\nu_j}_{LL}+C^{Z',\nu_i\nu_j}_{RL}}{C^{\text{SM},\nu_i\nu_i}_{LL}}\right|^2 \label{bsnunu}.
\ee 
For the decay $B^+\to K^+\bar{\nu}\nu$, the SM predicts BR$(B^+\to K^+\bar{\nu}\nu)=5.58(37)\times 10^{-6}$, including long distance contribution by double-charged current interaction $B^+\to \tau^+ \nu_{\tau}\to K^+\bar{\nu}_{\tau}\nu_{\tau}$ \cite{Parrott:2022zte}. On the one hand, the very recent result of the Belle-II experimental has shown the first evidence for the observation of BR$(B^+\to K^+\bar{\nu}\nu)=2.3(7)\times 10^{-5}$ \cite{Belle-II:2023esi}. In contrast, for the decay $B\to K^{*0} \bar{\nu}\nu$, there has not been observed signal yet, instead we have its upper limit, namely BR$(B\to K^{*0}\bar{\nu}\nu)<2.7\times 10^{-5}$ \cite{Belle:2017oht}. These lead to the two constraints of Eq. (\ref{bsnunu}) as follows 
\be R_{K^+}^{\nu}=4.12\pm 1.28, \hs R_{K^*}^{\nu}<3.2. \label{quarkcon1 }\ee  
Enhanced by the obtained limits in Eq. (\ref{quarkcon}), the model numerically estimates the ratio $R_{K}^{\nu}\simeq [1.0025,1.0048]$ which satisfies $2.5\sigma$ range of $R_{K^+}^{\nu}$ constraint as well as upper limit of $R_{K^*}^{\nu}$. Therefore, besides the mentioned quark flavor observables, the considering model can also relax constraints of the newest Belle-II result for the branching ratio of $B^+\to K^+\bar{\nu}\nu$ and $B\to K^{*0}\bar{\nu}\nu$ decays.

\subsection{\label{cLFV}Charged-lepton flavor violation}

As previously mentioned, we work in the basis of charged lepton mass eigenstates, i.e., $V_{e_{L,R}}=1$, then $\Gamma^{e_{L,R}}_{i,j}=0$ for $i\neq j$, so the model under consideration does not predict cLFV interactions mediated by $Z'$. However, the Yukawa couplings $h_x$ and $k_a$ associated with the neutrino mass generation (see Fig. \ref{fig1}) can generate cLFV processes at one-loop level, such as $e_i\to e_j\gamma$ and $e_i\to e_j\bar{e}_je_j$ ($3e_j$) decays with $e_{i,j}=e,\mu,\tau$ for $i,j=1,2,3$, and $\mu-e$ conversion in nuclei. In addition, the decays $\tau\to\mu\gamma$ and $\tau\to 3\mu$ receive contributions from both the seesaw and scotogenic mechanisms, whereas the decays $\tau/\mu\to e\gamma$ and $\tau/\mu\to 3e$ and the $\mu-e$ conversion only include to the scotogenic contribution.

For the radiate lepton decays $e_i\to e_j\gamma$, their branching ratios (BRs) are approximately given by \cite{Ilakovac:1994kj,Toma:2013zsa,Dinh:2013vya,Hagedorn:2018spx}
\bea 
\text{BR}(\mu\to e\gamma) &=& \frac{48\pi^3\al_\text{em}}{G_F^2}|G^{21}_\text{D}|^2\text{BR}(\mu\to e\nu_\mu\bar{\nu}_e),\\
\text{BR}(\tau\to e\gamma) &=& \frac{48\pi^3\al_\text{em}}{G_F^2}|G^{31}_\text{D}|^2\text{BR}(\tau\to e\nu_\tau\bar{\nu}_e),\\
\text{BR}(\tau\to \mu\gamma) &=& \left(\frac{48\pi^3\al_\text{em}}{G_F^2}|G^{32}_\text{D}|^2+\frac{3\al_\text{em}}{32\pi}\frac{v^4}{M_\nu^4}h_2^2h_3^2\left|\mathcal{S}(u_s)\right|^2\right)\text{BR}(\tau\to \mu\nu_\tau\bar{\nu}_\mu),\label{taumuga} \eea
where the parameter $u_s$ is defined by $u_s = M_\nu^2/m_W^2$. The dipole form factors $G_\text{D}^{x1,32}$ for $x=2,3$ are related to the scotogenic contribution and are defined as
\bea G^{x1}_\text{D}&=&\frac{s_{2\theta} k_xk_1}{64\pi^2}\left(\frac{1}{m^2_{H_1^\pm}}\mathcal{G}(u_1)-\frac{1}{m^2_{H_2^\pm}}\mathcal{G}(u_2)\right),\\
G^{32}_\text{D} &=&-\frac{k_3k_2}{32\pi^2} \left(\frac{s^2_\theta}{m^2_{H_1^\pm}}\mathcal{G}(u_1)+\frac{c^2_\theta}{m^2_{H_2^\pm}}\mathcal{G}(u_2)\right),\eea
with the parameters $u_{1,2} = M_N^2/m^2_{H_{1,2}^\pm}<1$. Above, the loop functions have the form:
\be \mathcal{G}(u)=\frac{1-6u+3u^2+2u^3-6u^2\ln u}{6(1-u)^4}, \hs \mathcal{S}(u_s)=3u_s\mathcal{G}(u_s). \ee
Additionally, it is well known that $\text{BR}(\mu\to e\nu_\mu\bar{\nu}_e)\simeq 1$, $\text{BR}(\tau\to e\nu_\tau\bar{\nu}_e)\simeq 0.178$, and $\text{BR}(\tau\to \mu\nu_\tau\bar{\nu}_\mu)\simeq 0.174$ \cite{ParticleDataGroup:2022pth}.

For the leptonic decays $e_i\to 3e_j$, there are four types of one-loop diagrams to be $\gamma$-penguin, $Z$-penguin, $H$-penguin, and box diagram, in both the scotogenic and seesaw contributions \cite{Ilakovac:1994kj,Toma:2013zsa,Dinh:2013vya,Hagedorn:2018spx}. However, the contribution of the $H$-penguin diagrams is suppressed by the smallness of the involved Yukawa couplings. Additionally, the leading order scotogenic contribution of $Z$-penguin diagrams is proportional to the square of the charged lepton masses and thus negligible compared to the scotogenic contribution of the $\gamma$-penguin and box diagrams, see Ref. \cite{Toma:2013zsa} for more details. Therefore, the branching ratios for these processes can be calculated as \cite{Ilakovac:1994kj,Toma:2013zsa,Dinh:2013vya,Hagedorn:2018spx}
\bea
\text{BR}(\mu\to 3e) &=& \frac{6\pi^2\al_\text{em}^2}{G_F^2}\left\{|H^{21}_\text{ND}|^2+|G^{21}_\text{D}|^2\left(\frac{16}{3}\ln\frac{m_\mu}{m_e}-\frac{22}{3}\right)\right.\crn
&&\left.+\fr 1 6 |B^{21}|^2-4\left[H^{21}_\text{ND}G^{21}_\text{D}-\fr 1 6 (H^{21}_\text{ND}-2G^{21}_\text{D})B^{21}\right]\right\} \text{BR}(\mu\to e\nu_\mu\bar{\nu}_e),\\
\text{BR}(\tau\to 3e) &=& \frac{6\pi^2\al_\text{em}^2}{G_F^2}\left\{|H^{31}_\text{ND}|^2+|G^{31}_\text{D}|^2\left(\frac{16}{3}\ln\frac{m_\tau}{m_e}-\frac{22}{3}\right)\right.\crn
&&\left.+\fr 1 6 |B^{31}|^2-4\left[H^{31}_\text{ND}G^{31}_\text{D}-\fr 1 6 (H^{31}_\text{ND}-2G^{31}_\text{D})B^{31}\right]\right\} \text{BR}(\tau\to e\nu_\tau\bar{\nu}_e),\\
\text{BR}(\tau\to 3\mu) &=& \frac{6\pi^2\al_\text{em}^2}{G_F^2}\left\{|H^{32}_\text{ND}|^2+|G^{32}_\text{D}|^2\left(\frac{16}{3}\ln\frac{m_\tau}{m_\mu}-\frac{22}{3}\right)\right.\crn
&&\left.+\fr 1 6 |B^{32}|^2-4\left[H^{32}_\text{ND}G^{32}_\text{D}-\fr 1 6 (H^{32}_\text{ND}-2G^{32}_\text{D})B^{32}\right]\right\} \text{BR}(\tau\to \mu\nu_\tau\bar{\nu}_\mu)\crn
&&+\frac{\al^2_\text{em}}{32\pi^2s_W^4}\frac{v^4}{M_\nu^4}h_2^2h_3^2\left\{|\mathcal{S}_\text{Box}(u_s)+c_{2W}\mathcal{S}_Z(u_s)+2s^2_W\mathcal{S}_\gamma(u_s)|^2+2s_W^4|\mathcal{S}_Z(u_s)-\mathcal{S}_\gamma(u_s)|^2\right.\crn
&&+2s_W^2[\mathcal{S}_\text{Box}(u_s)+\mathcal{S}_Z(u_s)]\mathcal{S}(u_s)-6s_W^4[\mathcal{S}_Z(u_s)-\mathcal{S}_\gamma(u_s)]\mathcal{S}(u_s)\crn
&&\left.+s_W^4|\mathcal{S}(u_s)|^2\left(\ln\frac{m^2_\tau}{m^2_\mu}-\frac{11}{4}\right)\right\}\text{BR}(\tau\to\mu\nu_\tau\bar{\nu}_\mu),\eea
where the non-dipole form factors $H^{x1,32}_\text{ND}$ are generated by non-dipole photon penguins, whereas the form factors $B^{x1,32}$ are induced by box-type diagrams, which have the form:
\bea H^{x1}_\text{ND} &=&\fr 1 3 G^{x1}_\text{D}|_{\mathcal{G}(u)\to\mathcal{H}(u)},\hs  H^{32}_\text{ND}=\fr 1 3 G^{32}_\text{D}|_{\mathcal{G}(u)\to\mathcal{H}(u)},\\
B^{x1} &=&-\frac{s_{2\theta}k_xk_1^3}{128\pi^3\al_\text{em}}\left(\frac{c^2_{\theta}}{m^2_{H_1^\pm}}\mathcal{B}(u_1)-\frac{s^2_{\theta}}{m^2_{H_2^\pm}}\mathcal{B}(u_2)-\frac{2c_{2\theta}}{m^2_N}\mathcal{B}(u_1,u_2)\right),\\
B^{32} &=& \frac{k_3k_2^3}{64\pi^3\al_\text{em}}\left(\frac{s^4_{\theta}}{m^2_{H_1^\pm}}\mathcal{B}(u_1)+\frac{c^4_{\theta}}{m^2_{H_2^\pm}}\mathcal{B}(u_2)+\frac{s_{2\theta}^2}{m^2_N}\mathcal{B}(u_1,u_2)\right). \eea
Above, the loop functions are defined by
\bea \mathcal{H}(u) &=&\frac{2-9u+18u^2-11u^3+6u^3\ln u}{6(1-u)^4}, \hs \mathcal{B}(u)=\frac{1-u^2+2u\ln u}{2(u-1)^3},\\
\mathcal{B}(u_1,u_2) &=& -\frac{u_1u_2}{2(1-u_1)(1-u_2)}-\frac{u_1^2u_2\ln u_1}{2(1-u_1)^2(u_1-u_2)}-\frac{u_1u_2^2\ln u_2}{2(1-u_2)^2(u_2-u_1)},\\
\mathcal{S}_\text{Box}(u_s) &=&-\frac{u_s}{2(1-u_s)}\left(1+\frac{\ln u_s}{1-u_s}\right),\hs \mathcal{S}_Z(u_s)=-\frac{u_s}{4(1-u_s)}\left(5+\frac{2+3u_s}{1-u_s}\ln u_s\right),\\
\mathcal{S}_\gamma(u_s)&=&\frac{u_s(7u_s^2-u_s-12)}{24(1-u_s)^3}-\frac{u_s^2(12-10u_s+u_s^2)}{12(1-u_s)^4}\ln u_s. \eea

Next, we consider $\mu-e$ conversion in nuclei based on the framework of our model. This process receives contributions from $\gamma$-, $Z$- and $H$-penguin diagrams \cite{Toma:2013zsa}. However, the leading order contribution from the $Z$-penguins is proportional to the square of the charged lepton masses and thus negligible compared to the $\gamma$-penguin contribution. Additionally, the $H$-penguin contribution is suppressed by the smallness of the involved Yukawa couplings. Therefore, the $\mu-e$ conversion is dominated only by the $\gamma$-penguin diagrams. Further, we concentrate on the coherent conversion processes in which the final state of the nucleus $\mathcal{N}$ is the same as the initial one, and then the matrix elements of $\langle\mathcal{N}|\bar{q}\gamma_\mu\gamma_5 q|\mathcal{N}\rangle$, $\langle\mathcal{N}|\bar{q}\gamma_5 q|\mathcal{N}\rangle$, and $\langle\mathcal{N}|\bar{q}\sigma_{\mu\nu} q|\mathcal{N}\rangle$ vanish identically. In this case, the $\mu-e$ conversion rate (CR) in a target of atomic nuclei, relative to the muon capture rate, can be calculated as \cite{Kuno:1999jp,Arganda:2007jw}
\be \text{CR}(\mu-e,\mathcal{N})=\fr{p_eE_em_\mu^3G_F^2\al^3_\text{em}Z_\text{eff}^4F_p^2}{8\pi^2 Z\Gamma_\text{capt}}\left|Z\left(2g^\gamma_{LV(u)}+g^\gamma_{LV(d)}\right)+N\left(g^\gamma_{LV(u)}+2g^\gamma_{LV(d)}\right)\right|^2, \label{mue}\ee
where $Z$ and $N$ are the number of protons and neutrons in the nucleus, $Z_\text{eff}$ is the effective atomic charge, $F_p$ is the nuclear matrix element, and $\Gamma_\text{capt}$ is the total muon capture rate. The numerical values of these parameters for the nuclei used in current or near future experiments can be found in \cite{Arganda:2007jw} and references therein, such as $[Z,N,Z_\text{eff},F_p,\Gamma_\text{capt}\times 10^{18}(\text{GeV})]=[22,26,17.6,0.54,1.70422]$, $[79,118,33.5,0.16,8.59868]$, and $[82,125,34.0,0.15,8.84868]$, for titanium (Ti), gold (Au), and lead (Pb), respectively. In addition, $p_e$ and $E_e$ are the momentum and energy of the electron, which are set to $m_\mu$ in the numerical evaluation. The effective couplings $g_{LV(q)}^\gamma$ for $q=u,d$ are associate with a left-handed leptonic vector current, generated by photon penguins,
\be g_{LV(q)}^\gamma=\fr{\sqrt2}{G_F}e^2Q_q(H^{21}_\text{ND}-G^{21}_\text{D}),\ee
 with $Q_q$ to be the electric charge of the quark $q$. We note that in Eq. (\ref{mue}) a right-handed leptonic vector current is not generated at one-loop level, i.e., $g_{RV(q)}^\gamma=0$, because the new scalar doublets $\eta$ and $\rho$ only couple to the left-handed lepton doublets $l_{1L}$ and $l_{xL}$, respectively.
 
\begin{table}[h!]
	\begin{centering}
		\begin{tabular}{|c|c|c|}
\hline cLFV process & Current bound & Future sensitivity  
\tabularnewline \hline 
			BR$ (\mu \to e\ga)$ & $3.1\times 10^{-13}$  \cite{MEGII:2023ltw} & $6.0\times 10^{-14}$ \cite{MEGII:2018kmf} 
\tabularnewline
			BR$(\tau\to e\ga)$ & $3.3\times 10^{-8}$ \cite{BaBar:2009hkt} & $3.0\times 10^{-9}$ \cite{Belle-II:2018jsg}
\tabularnewline
			BR$ (\tau\to \mu\ga)$ & $4.4\times 10^{-8}$  \cite{BaBar:2009hkt} & $1.0\times 10^{-9}$ \cite{Belle-II:2018jsg}
\tabularnewline 
			BR$ (\mu\to 3e)$ & $1.0\times 10^{-12}$  \cite{SINDRUM:1987nra} & $1.0\times 10^{-16}$ \cite{Blondel:2013ia}
\tabularnewline
			BR$(\tau\to 3e)$ & $2.7\times 10^{-8} $ \cite{Hayasaka:2010np} & $5.0\times 10^{-10}$ \cite{Belle-II:2018jsg}
\tabularnewline
			BR$(\tau\to 3\mu)$ & $2.1\times 10^{-8} $ \cite{Hayasaka:2010np} & $4.0\times 10^{-10}$ \cite{Belle-II:2018jsg}
\tabularnewline
			CR($\mu$ -- $e$, Ti)& $4.3\times 10^{-12}$ \cite{SINDRUMII:1993gxf} & $1.0\times 10^{-18}$ \cite{Alekou:2013eta}
\tabularnewline
			CR($\mu$ -- $e$, Au)& $7.0\times 10^{-13}$ \cite{SINDRUMII:2006dvw} & No data
\tabularnewline
			CR($\mu$ -- $e$, Pb)& $4.6\times 10^{-11}$ \cite{SINDRUMII:1996fti} & No data
\tabularnewline\hline 
		\end{tabular}
	\par
	\protect\caption{\label{leptonfalvor} Current experimental bounds and future sensitivities for the BRs and CRs of cLFV processes.}
	\end{centering}
\end{table}

\begin{figure}[h]
\centering
\includegraphics[scale=0.42]{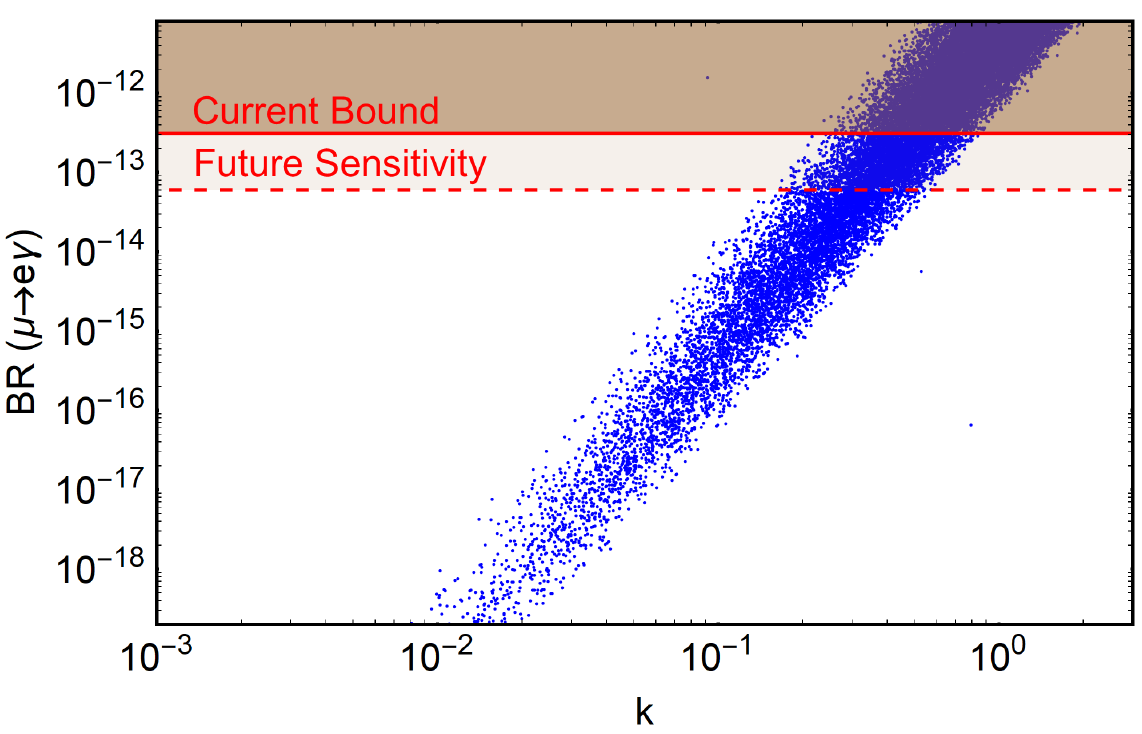}
\includegraphics[scale=0.42]{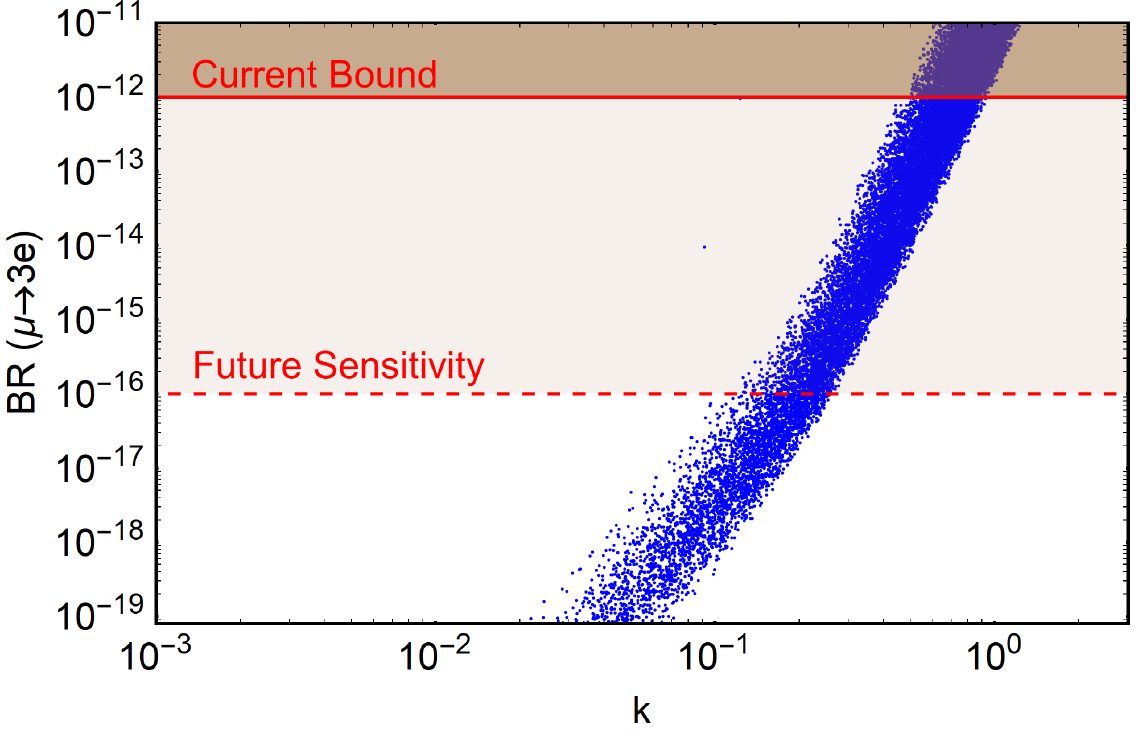}
\includegraphics[scale=0.42]{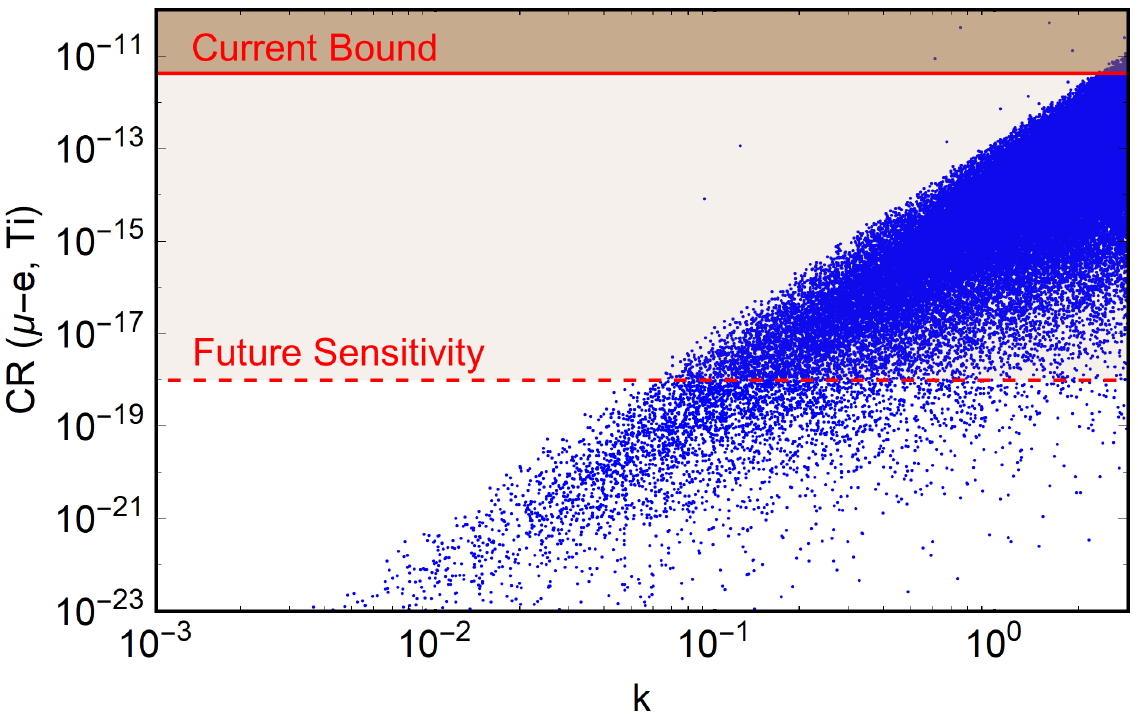}
 \caption[]{\label{fig5}BRs and CR as a function of the magnitude of Yukawa coupling $k$. The solid and dashed red lines indicate the current bounds and future sensitivities, respectively. The parameter space according to shaded regions is excluded.}
\end{figure}

In Table \ref{leptonfalvor} we present the current experimental limits and future sensitivities for the BRs and CRs of cLFV processes. First of all, we can estimate that the seesaw contributions are negligible, $\sim 10^{-31}$ for $\text{BR}(\tau\to \mu\gamma)$ and $\sim 10^{-32}$ for $\text{BR}(\tau\to 3\mu)$, compared to their experimental bounds, $\sim 10^{-8}$ \cite{BaBar:2009hkt,Hayasaka:2010np}, taking $h_{2,3}\sim 10^{-5.9}$ and $M_\nu\sim 1$ TeV, as implied from the neutrino mass generation. Regarding the scotogenic contributions to these BRs, they are strongly suppressed by the largeness of charged odd-scalar masses and/or the smallness of the Yukawa couplings. For example, take $m_{H_{1,2}^\pm}\sim m_N\sim 1$ TeV, the current experimental bounds and expected future sensitivities on these BRs can be satisfied if $k_{2,3}\lesssim 1$. Whereas, if $m_{H_{1,2}^\pm}\sim m_N\sim 0.1$ TeV, the maximum allowed Yukawa coupling must be lowered to $0.1$.\footnote{The lower bound on charged scalar mass, imposed by LEP, is in the range [70, 90] GeV \cite{ALEPH:2013htx}.} For the remaining decays, the scotogenic contribution is suppressed by not only the largeness of charged odd-scalar masses and/or the smallness of the Yukawa couplings but also either the small mixing among $\eta^\pm$ and $\rho^\pm$ or the degeneracy among charged odd-scalar masses. Since the present work is considering the mixing among $\eta^\pm$ and $\rho^\pm$ to be small, it is natural to assume that the charged odd-scalar masses are not degenerate. In Fig. \ref{fig5}, we present the strongest constraints for the magnitude of Yukawa couplings $k_{1,2,3}$, generally denoted $k$, which come from the cLFV processes considered above. Here, we randomly seed the free parameters in ranges as $k\in [10^{-3},\sqrt{4\pi}]$, $m_{N,H_{1,2}^\pm}\in [0.5,1.2]$ TeV, and require the small mixing, i.e., $|\theta|\sim 0.5\times\arctan(v^2/|m^2_{H_2^\pm}-m^2_{H_1^\pm}|)<\pi/16$. This figure show that the current experimental bounds, indicated by the solid red lines, constrain $k\lesssim 0.8$. Further, the future sensitivities for these processes, indicated by the dashed red lines, imply the upper bound of $k$ must be lowered to $0.2$. 

\section{\label{dm}Dark matter}
Our model contains two potential candidate kinds for DM: dark singlet fermion ($N_R$) and dark doublet scalar ($R_{1,2}, I_{1,2}$). This section focuses on studying the viable DM candidate, either $N_R$ or $R_1$, which is assumed to be the lightest of the dark fields. Additionally, in the limit $\La_{1,2}\gg v,\mu$ then the mixings in the scalar sector are tiny, i.e., $\ep_{1,2}, \theta_{R,I},\theta\ll 1$, thus these mixings can be safely neglected in the calculation here. In addition, this section assumes $\La_1\simeq \La_2\equiv \La$ for simplicity.

\subsection{Fermion dark matter}
Assuming that $N_R$ is the lightest among the dark particles, i.e., $M_N<m_{R_{1,2}}, m_{I_{1,2}}, m_{H^\pm_{1,2}}$, thus $N_R$ is absolutely stabilized by the residual matter parity $\mathbb{Z}_2$ and responsible for DM. Since $N_R$ is a gauge singlet, $N_R\sim (1,1,0,0)$, its production mechanism in the early Universe depends on the magnitude of Yukawa couplings $k_a$, which couple $N_R$ with SM leptons and the dark doublet scalars, $\eta$ and $\rho$. As mentioned in the last paragraph of Section \ref{model}.D., the scotogenic contribution to neutrino mass is appropriate to the experiment if $k\la_{19} \sim 10^{-4}$ and $M_N\sim 1$ TeV, where $k\sim k_a$ for $a=1,2,3$. Hence, if $\la_{19}$ is made small, then $k$ can be enhanced and vice versa, and it is not possible to obtain $k$ smaller than $\mathcal{O}(10^{-4})$ even if $\la_{19}$ is $\mathcal{O}(1)$. In other words, the viable DM candidate $N_R$ is significantly coupled to the normal matter in the thermal bath of the early Universe. Note that the dark doublet scalars are always in thermal equilibrium with the SM plasma since they are coupled to the Higgs and gauge portals via the couplings $\la_{11,12,13,14}$ and/or $g_X$. Hence, the freeze-out mechanism works, determining the DM relic density and implying the DM's nature as a weakly interacting massive particle (WIMP).\footnote{Suppose the particle content is added by a singlet fermion $N'_R\sim (1,1,0,0)$ that is lighter than $N_R$ and has Yukawa coupling to be very small. In that case, $N'_R$ is a viable DM candidate and gives a negligible contribution to neutrino masses. Additionally, $N'_R$ does not reach thermal equilibrium in the early Universe and may be produced via a freeze-in mechanism, where its relic density is dominated by the decays of the odd scalars.} 

The computation of the DM relic density $\Om_\text{DM} h^2$ and DM-nucleon spin-independent (SI) cross section at tree level $\sigma^\text{SI}$ is performed with the micrOMEGAs-6.0.5 \cite{Alguero:2023zol,Belanger:2018ccd}. Regarding input parameters, we randomly seed $k$ and $m_{N_R}$ in ranges as $k\in [10^{-3},\sqrt{4\pi}]$ and $m_{N_R}\in [0.5,1.2]$ TeV, which are similar in cLFV studies. The mass of dark charged Higgs bosons $H_{1,2}^\pm$ is also seeded in the range $[0.5,1.2]$ TeV, but with additional constraint  $m^2_{H_2^{\pm}}>v^2/\tan{(\pi/8)}+m^2_{H_1^{\pm}}$, also inspired by cLFV studies. They suggest that the DM relic density can be mostly governed by the (co)annihilation of $H_1^{\pm}, R_1, I_1$, whereas the fields $H_2^{\pm}, R_2, I_2$ contribute insignificantly. Otherwise, the coupling $g_X$ is chosen to satisfy quark flavor studies, i.e., $|z|g_X\gtrsim 0.059$; for instance, we set $|z|=1$ and $g_X=0.08$. In addition, the Higgs scalar coupling $\la_{19}$ is modified by the values of $k$ and $m_{N_R}$ due to the following constraint obtained by loop contribution to active neutrino mass, $k\la_{19}\sim 10^{-4} \ [\text{TeV}^{1/2}]/\sqrt{m_{N_R}}$. It is worth stressing that if $m_{H_{1,2}}/2 \in [0.5,1.2]$ TeV, there are two resonances in relic density, since the (co)annihilation of $H_1^{\pm}, R_1, I_1$ through $s$-channel $H_{1,2}$ mediation. Therefore, the Higgs couplings $\la_{2,3,6}$ and $\la$ which affect to $m_{H_{1,2}}$ are chosen such that either $m_{H_{1,2}}/2\in [0.5,1.2]$ TeV or $m_{H_{1,2}}/2 > 1.2$ TeV. We also assume that $\la_{11,12,13,14}\sim \mathcal{O}(10^{-2}-10^{-1})$.\footnote{It is checked that if we choose larger values for the Higgs couplings, i.e., $\la_{11,12,13,14}\sim\mathcal{O}(10^{-1}-1)$, then the viable parameter space for fermionic DM scenario is relaxed. However, the scalar DM scenario is ruled out since it cannot account for the observed DM relic density.}  

\begin{figure}[h!]
\bc
\includegraphics[scale=0.30]{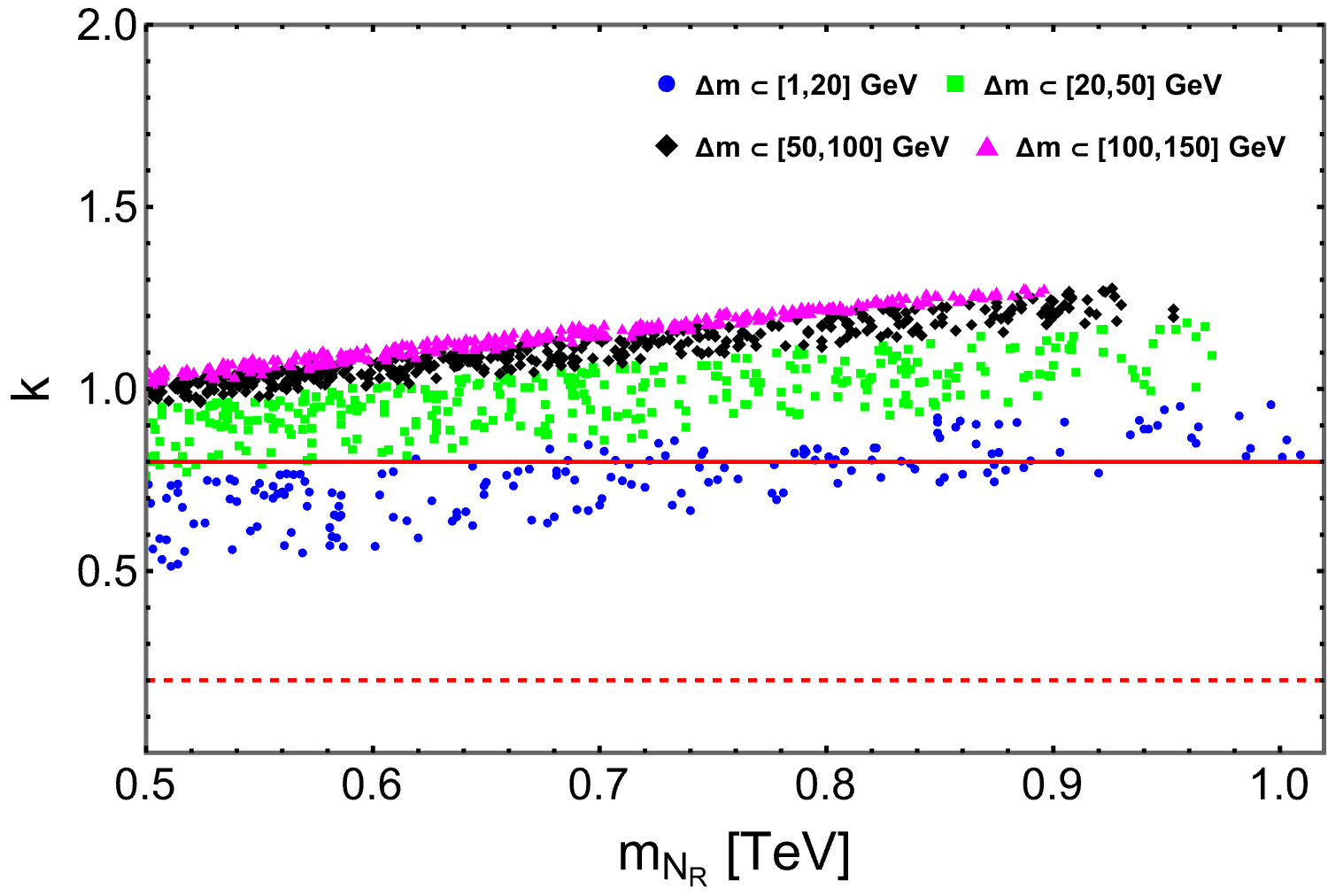} 
\includegraphics[scale=0.30]{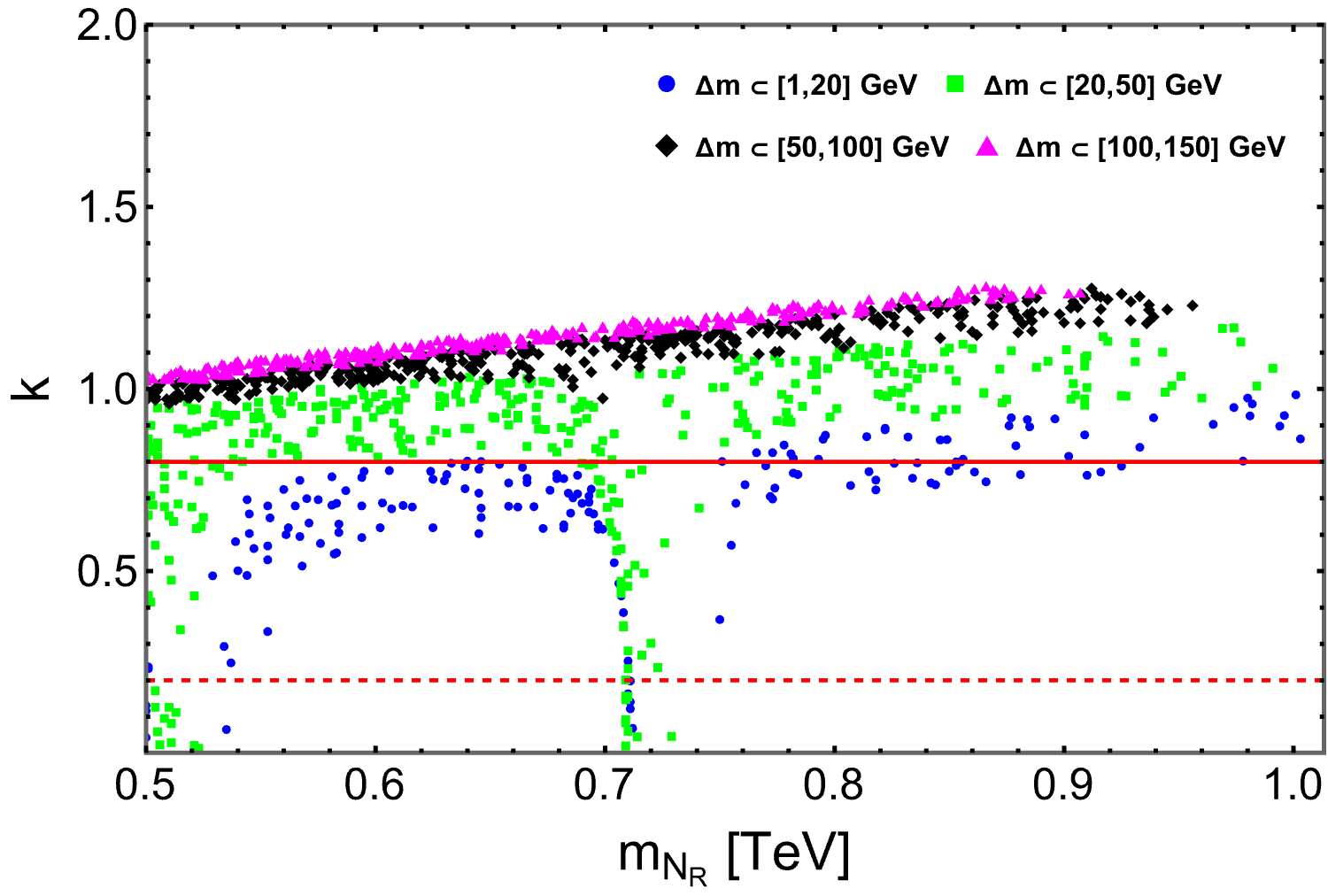}
\caption[]{The correlations between Yukawa coupling $k$ and mass of fermionic DM $m_{N_R}$ satisfying the $3\sigma$ range of CMB observation for DM relic density \cite{Planck:2018nkj} with $m_{H_{1,2}}/2> 1.2$ TeV (left panel) and $m_{H_{1,2}}/2\in [0.5,1.2]$ TeV (right panel). The parameter $\Delta m\equiv m_{H_1^{\pm},R_1,I_1}-m_{N_R}$ denotes the degeneracy in mass. The solid and dashed red lines represent the upper limits of $k$ associated with the cLFV processes, $k\lesssim 0.8$ (current bound) and $k\lesssim 0.2$ (future sensitivity).}
\label{fig6}
\ec 
\end{figure}

\begin{figure}[h!]
\centering
\begin{subfigure}[b]{0.9\textwidth}
\centering
\includegraphics[scale=1.4]{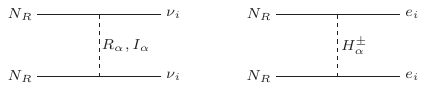}
\caption[]{Annihilations of fermion DM into SM leptons.}
\label{fig7a}
\end{subfigure}

\begin{subfigure}[b]{0.9\textwidth}
\centering
\includegraphics[scale=1.4]{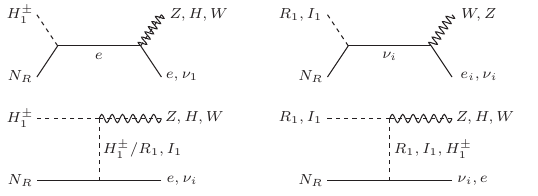}
\caption[]{Coannihilations of fermion DM with dark scalar particles.}
\label{fig7b}
\end{subfigure}

\begin{subfigure}[b]{0.9\textwidth}
\centering
\includegraphics[scale=1.4]{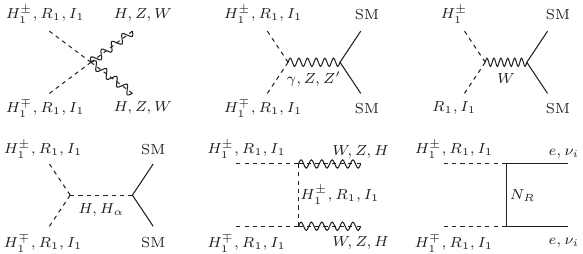}
\caption[]{Annihilations of coannihilation partners.}
\label{fig7c}
\end{subfigure}

\caption[]{Annihilations and coannihilations govern the DM relic density, where $\al=1,2$.}
\label{fig7}
\end{figure}

The left panel of Fig. \ref{fig6} corresponds to case $m_{H_{1,2}}/2 > 1.2$ TeV, in which the correlations between Yukawa coupling $k$ and fermionic DM mass $m_{N_R}$ fulfilling 3$\sigma$ the constraint of relic density for DM by CMB observation $\Om_{\text{DM}}h^2=0.11933\pm 0.00091$ \cite{Planck:2018nkj}, are demonstrated. Here, the correlations are plotted for each mass difference between dark particles $H_{1}^{\pm},R_1,I_1$ and DM candidate $N_R$, $\Delta m=m_{H_1^{\pm},R_1,I_1}-m_{N_R}$. We firstly focus on the cases $\Delta m\in [1,20] $ GeV and $\Delta m\in [20,50]$ GeV, which imply there are nearly degeneracy between $m_{N_R}$ and $m_{H_{1}^{\pm},R_1,I_1}$. Hence, the relic abundance of the lightest dark particle $N_R$ is determined by not only its annihilation but also the annihilation of particles like $H_1^{\pm},R_1,I_1$. The diagrams in subfigure \ref{fig7b} are coannihilations of $N_R$ with dark scalar particles into a lepton and a gauge boson or Higgs boson. The annihilation of the coannihilation partners directly into Higgs bosons, via gauge interactions into SM particles, or via DM-mediated $t$-channel into leptons, is described by diagrams in subfigure \ref{fig7c}. We see that the values of Yukawa coupling $k$ not only tend to increase with $m_{N_R}$ but also with $\Delta m$. For instance $k$ varies in the following ranges $[0.52,0.973]$, $[0.772,1.17]$, $[0.97,1.272]$ and $[1.02,1.281]$ for $\Delta m\in [1,20]$ GeV, $\Delta m\in [20,50]$ GeV, $\Delta m\in [50,100]$ GeV and $\Delta m\in [100,150]$ GeV, respectively. However, to satisfy the current constraint from cLFV studies, $k\lesssim 0.8$ (shown as the solid red line), the only scenario with lowest mass degeneration $\Delta m\in [1,20]$ GeV is shown to be adapted, but cannot explain the future sensitivity $k\lesssim 0.2$ (denoted as dashed red line). In particular, this case satisfies the current cLFV constraint $k\lesssim 0.8$ only if $m_{N_R}\in[0.5,0.979]$ TeV. Thus, the model within the mass degeneration between fermionic DM $m_{N_R}$ and other dark particles $\Delta m\in [1,20]$ GeV, the viable range of $k$ and $m_{N_R}$ satisfying both constraints of relic density \cite{Planck:2018nkj} and current cLFV studies are $k\in[0.52,0.8]$ and $m_{N_R}\in[0.5,0.979]$ TeV, but does not reach the future cLFV sensitivity $k\lesssim 0.2$. 

Moving to the right panel of  Fig. \ref{fig6} where $m_{H_{1,2}}/2\in [0.5,1.2]$ TeV, we see that the behaviors in the low mass degeneracy cases $\Delta m\in [1,20]$ GeV and $\Delta m\in [20,50]$ GeV change remarkably, in compared to corresponding ones in the left panel. We find two distinct resonances according to $m_{N_R}\sim m_{H_{1,2}}/2$, which are relate to the (co)annihilation of $H_1^\pm,R_1,I_1$ via $s$-channel with $H_{1,2}$ to be mediators such as $H_1^+H_1^-, R_1R_1, I_1I_1\to W^+W^-,HH$. Indeed, when fermionic DM mass $m_{N_R}$ attains around a half of mediators $m_{H_{1,2}}/2$, the relic density usually turns to be very suppressed since the annihilation cross section is proportional $1/(m_{H_{1,2}}^2-4m_{N_R}^2)$. However, the other contributions, such as the diagrams of subfigure \ref{fig7b} and the last one in subfigure \ref{fig7c}, trigger relic abundance related to $1/k^2$. Thus, if $k$ is small enough, these contributions will give large values for relic density, and then the $\Omega_{N_R} h^2$ can attain desirable values. Moreover, the case $\Delta m\in [20,50]$ GeV gives correlated points satisfying cLFV constraint $k\lesssim 0.8$, in contrast to the left panel. It is worth noting that both scenarios $\Delta m\in [1,20]$ GeV and $\Delta m\in [20,50]$ GeV even adapt the future cLFV sensitivity $k\lesssim 0.2$. Consequently, the right panel implies a better explanation for the current bound and future sensitivity of cLFV processes besides fulfilling CMB observation for DM relic abundance \cite{Planck:2018nkj}.

\begin{figure}[h!]
\bc
\includegraphics[scale=1]{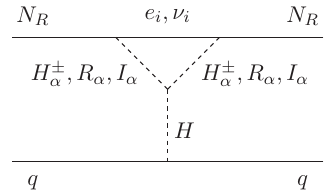}
\caption[]{\label{fig8}Diagrams for DM-nucleon scattering, where $\al=1,2$.}
\ec
\end{figure}

Next, we discuss the direct detection searches for the fermion DM $N_R$. Although there are no direct interactions between $N_R$ with SM quarks at the tree level, $N_R$ can have effective couplings with various SM particles, such as the photon, $Z$ boson, and Higgs boson, at the one-loop level. Additionally, the $Z$-boson exchange leads to an effective axial-vector interaction, which gives rise to a spin-dependent DM-nucleon scattering and is dominant if the couplings between Higgs and the dark scalars are very small. Since the constraint of spin-dependent DM-nucleon scattering is relatively less constrained than the SI scattering, we focus here on the SI DM-nucleon scattering exchanged by the Higgs boson, as described in Fig. \ref{fig8}. 

The effective Lagrangian describing the scattering of $N_R$ with quarks is given by\be\mathcal{L}_{\mathrm{eff}}= C_\mathrm{SI}m_q(\bar{N}_R^cN_R)(\bar{q}q),\ee 
where the effective coupling of $N_R$ to quarks are calculated as
\bea C_\mathrm{SI} &\simeq& -\frac{k^2}{16\pi^2m_H^2M_N}\left(\la_{11}\mathcal{I}(m_{H_1^\pm}^2,M_N^2)+\la_{12}\mathcal{I}(m_{H_2^\pm}^2,M_N^2)\right.\crn
&&\left.+\fr{\la_{11}+\la_{13}}{2}\mathcal{I}(M_\eta^2,M_N^2)+\fr{\la_{12}+\la_{14}}{2}\mathcal{I}(M_\rho^2,M_N^2)\right) \eea
with loop function is calculated in the limit $m_{e_i,\nu_i}/m_{H^{\pm}_{\al}/R_{\al},I_{\al}}\ll 1$, given by the following expression $\mathcal{I}(a,b)=1-\fr{a-b}{b}\ln\frac{a}{a-b}$ for $a>b$. Then, the SI cross section for the interaction of $N_R$ with a proton is given by
\be \sigma^p_{\mathrm{SI}}=\frac{4m^2_p}{\pi}\left(\frac{M_Nm_p}{M_N+m_p}\right)^2C_\mathrm{SI}^2f_p^2, \ee 
where $f_p$ represents the scalar form factor, $f_p=0.326$ \cite{Mambrini:2011ik}. 

With the constraints obtained above, i.e., $k\lesssim 0. 8$, $m_{N_R}\in [0.5,0.9]$ TeV, and $\Delta m \in [1,20]$ GeV, we estimate the SI $N_R$-nucleon cross section $\sigma_{\text{SI}}^{N_R}\leq  \mathcal{O}(10^{-49}-10^{-48})$ cm$^2$. This predicted  $\sigma_{\text{SI}}^{N_R}$ is in agreement with the current upper experimental limits from XENONnT \cite{XENON:2023cxc}, LZ \cite{LZ:2022lsv}, and PandaX-4T \cite{PandaX-4T:2021bab}.

\subsection{Scalar dark matter}
We now turn our attention for the scalar DM scenario, which is assumed as $R_1$, i.e $m_{R_1}<m_{N_R},m_{R_2},m_{I_{1,2}}, m_{H_{1,2}^{\pm}}$. Since there are small mass splittings between DM candidate $R_1$ and other scalar dark particles $I_1$ and $H_1^\pm$ by the terms including $\la_{19}$, as shown in Eqs. (\ref{remass})--(\ref{chargedmass}), thus the relic abundance of $R_1$ is affected mostly by the coannihilation channels with $I_1$ and $H_1^{\pm}$. Similarly in the previous case of fermionic DM $N_R$, we also study the DM phenomenology in two separate cases, $m_{H_{1,2}}/2 > 1.2$ TeV and $m_{H_{1,2}}/2\in[0.5, 1.2]$ TeV. In addition, depending on input parameter, the scalar DM mass $m_{R_1}$ can be either degenerated with $m_{N_R}$, i.e. $m_{R_1}-m_{N_R}=\Delta m \in[1,20]$ GeV or non-degenerated with $m_{N_R}$.
 
\begin{figure}[h!]
\bc
\includegraphics[scale=0.30]{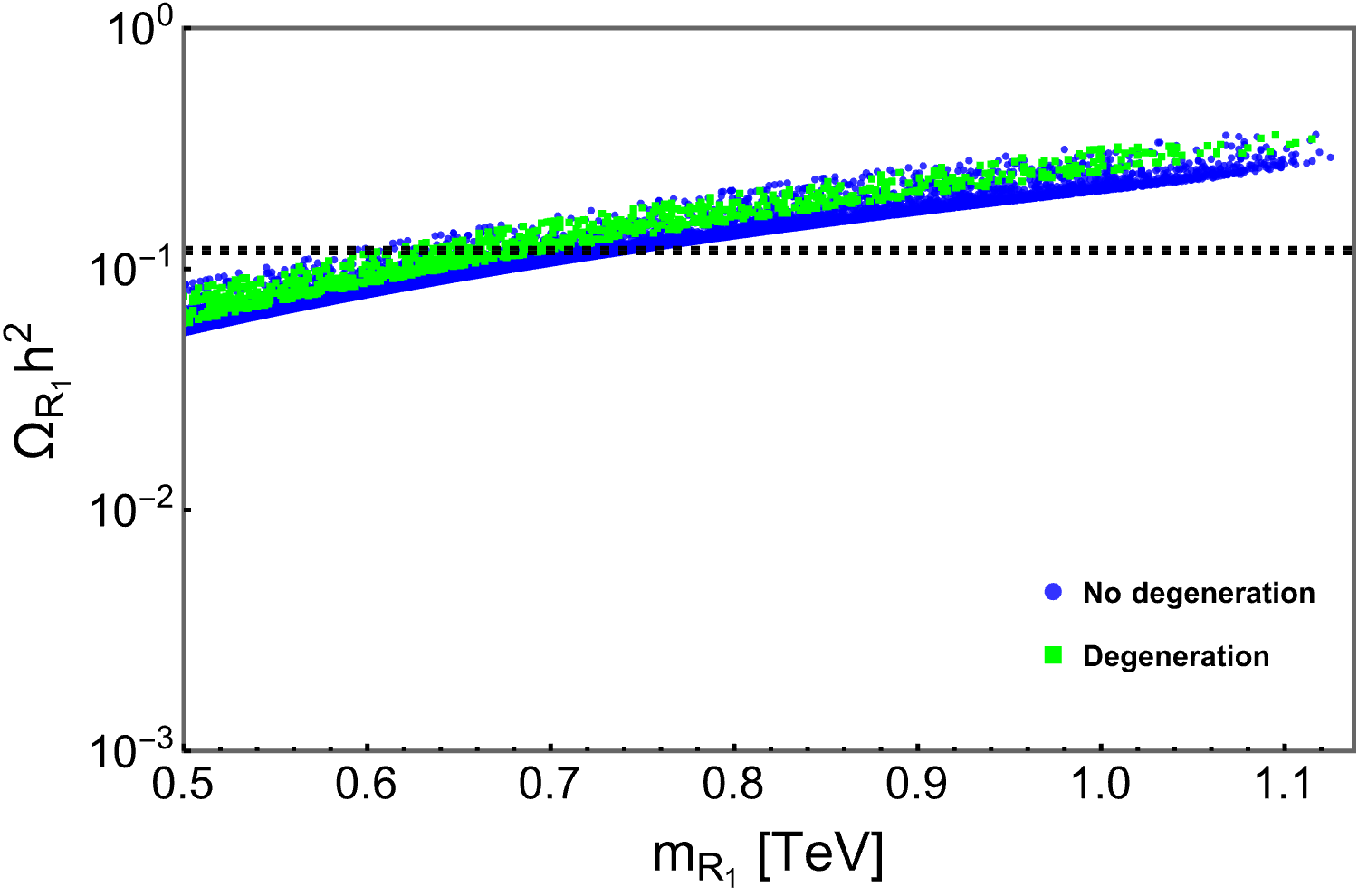} 
\includegraphics[scale=0.30]{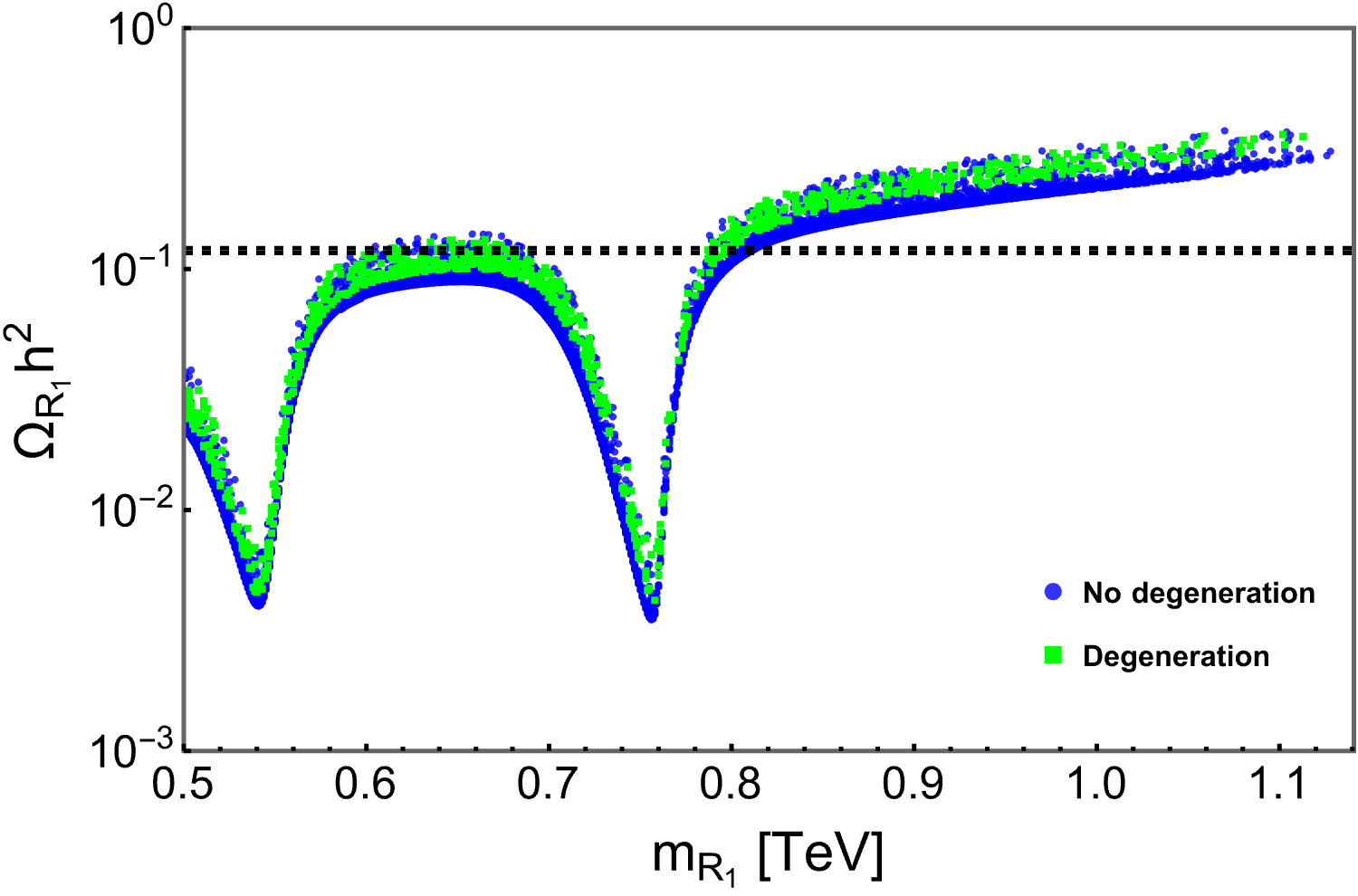}
\caption[]{\label{fig9}The relic abundance of scalar DM $R_1$ as a function of its mass $m_{R_1}$. Left panel is for the case $m_{H_{1,2}}/2>1.2$ TeV and right panel is for the case $m_{H_{1,2}}/2\in [0.5,1.2]$ TeV. The dot-dashed black lines represent the $3\sigma$ range of CMB observation \cite{Planck:2018nkj}. Both the panels are plotted in two cases, non-degeneration (blue points) and degeneration (green points) between $m_{R_1}$ and $m_{N_R}$.}
\ec
\end{figure}

In Fig. \ref{fig9}, we see that the left panel of the case $m_{H_{1,2}}/2>1.2$ TeV illustrates a different behavior in comparison to the right panel of the case $m_{H_{1,2}}/2\in [0.5,1.2]$ TeV. Indeed, in the left panel, the relic density increases almost linearly with $m_{R_1}$. In contrast, the relic density in the right panel depends more significantly on $m_{R_1}$ with two resonances according to $m_{R_1}\simeq m_{H_1}/2$ and $m_{R_1}\simeq m_{H_2}/2$. This behavior is similar to that of fermionic DM $N_R$. On the other hand, the relic density $\Omega_{R_1} h^2$ under the mass degeneration case between $m_{R_1}$ and $m_{N_R}$ is entirely overlapped on that of the non-degeneration case in both panels. This situation is different than Fig. \ref{fig6}, thus implying that the mass degeneration between $m_{R_1}$ and $m_{N_R}$ is not vital for the scalar DM scenario. Besides, we obtain the range of $m_{R_1}$ satisfying CMB observation \cite{Planck:2018nkj} is $m_{R_1}\in [0.6,0.71]$ TeV for left panel, whereas in the right panel, there are two satisfying distinct regions of $m_{R_1}$ which are $m_{R_1}\in [0.59,0.68]$ TeV and $m_{R_1}\in [0.793,0.798]$ TeV.

\begin{figure}[h!]
\bc
\includegraphics[scale=0.18]{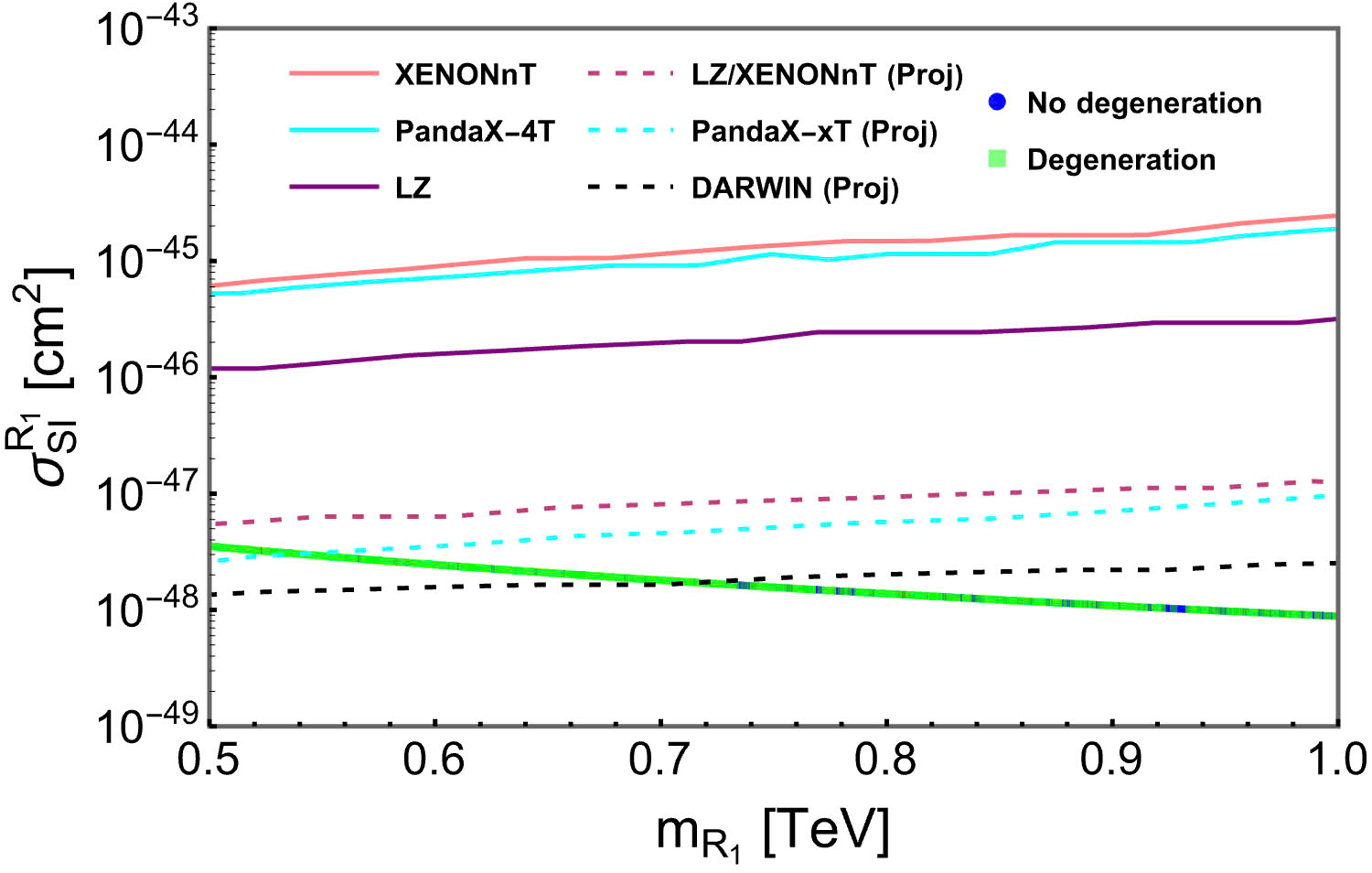} 
\caption[]{\label{fig10} The SI $R_1$-nucleon cross section $\sigma^{R_1}_\text{SI}$ as a function of $m_{R_1}$. The figure is plotted for two cases, non-degeneration (blue points) and degeneration (green points) between $m_{R_1}$ and $m_{N_R}$.}
\ec
\end{figure}

In contrast with the case of fermion DM $N_R$, the SI cross section for the scattering of scalar DM on nucleon $\sigma^{R_1}_\text{SI}$ can be induced exactly at the tree-level via exchange of Higgs bosons $H$ and $H_{1,2}$. It is important to note that there are also contributions from $Z$ exchange to this process since $R_1$ has non-zero hypercharge. However, as mentioned before, the masses of $R_1$ and $I_1$ are split by small term depending on $\la_{19}\neq 0$ (see Eqs. (\ref{remass})--(\ref{immass})), therefore the $\sigma^{R_1}_\text{SI}$ contributed by $Z$ will be kinematically forbidden, consequently making the Higgs boson effect becomes dominantly. The predicted $\sigma^{R_1}_\text{SI}$ is plotted as the function of $m_{R_1}$ in Fig. \ref{fig10}, where we have choose $m_{H_{1,2}}/2\in [0.5,1.2]$ TeV. We comment that the values of $\sigma^{R_1}_\text{SI}$ meet the upper experimental limits of XENONnT \cite{XENON:2023cxc}, LZ \cite{LZ:2022lsv} and PandaX-4T \cite{PandaX-4T:2021bab}, in which $\sigma^{R_1}_\text{SI}$ is lower than 2 to 3 orders in magnitude. In addition, the Fig. \ref{fig10} also presents the comparison of predicted $\sigma^{R_1}_\text{SI}$ with the future sensitivity projections including LZ/XENONnT \cite{LZ:2018qzl,XENON:2020kmp}, DARWIN \cite{DARWIN:2016hyl} and PandaX-xT \cite{PandaX:2024oxq}. This hints that the $\sigma^{R_1}_\text{SI}$ can satisfy the constraint of LZ/XENONnT with whole range of $m_{R_1}$, whereas for relaxing PandaX-xT and DARWIN limits, we obtain $m_{R_1}\geq 0.53$ TeV and $m_{R_1}\geq 0.72$ TeV.  
  
\section{\label{conclusion}Conclusion}
We have explored an extension of the SM, where the SM gauge symmetry is extended by the Abelian gauge symmetry group $U(1)_X$, for which the $X$ charge depends on flavors of quarks and leptons, in contrast to the usual charges like the electric charge and the hypercharge. We have shown that the proposed model can provide simultaneously possible solutions for several puzzles of the SM, including the observed fermion generation number, the neutrino mass generation mechanism, the existence of DM, and the flavor anomalies in both the quark and lepton sectors. The new physics effects of the new gauge boson $Z'$ at the LEP and LHC experiments have also been examined. 

We have determined the viable parameter space related to the additional gauge symmetry, which satisfies all the current experimental constraints, such as the mass of new gauge boson $m_{Z'}\gtrsim 4.64$ TeV, the product of charge parameter and coupling constant $|z|g_X\gtrsim 0.059$, the VEVs of new scalars $57.09\text{ TeV}\lesssim 2\sqrt{\La_1^2+9\La_2^2}\lesssim 78.92 \text{ TeV}$. Regarding DM, the model predicts two potential candidates, the dark Majorana fermion $N_R$ and the dark scalar $R_1$, with their masses, $M_N\sim 0.5$ -- 0.9 TeV and $M_{R_1}\sim 0.6$ -- 0.8 TeV. Besides, the magnitude of Yukawa couplings related to the neutrino mass generation is estimated as $h\sim 10^{-5.9}$ and $k\lesssim 0.8$ for the seesaw and scotogenic mechanisms, respectively. 

\section*{Acknowledgement}

This research is funded by Vietnam National Foundation for Science and Technology Development (NAFOSTED) under grant number 103.01-2023.50.

\bibliographystyle{JHEP}

\bibliography{combine}

\end{document}